\documentclass[
	amsmath,
	amssymb,
    aps,
    reprint,
    pre,
    showkeys,
]{revtex4-1}

\usepackage[utf8]{inputenc}
\usepackage{amsthm}
\usepackage{hyperref}
\usepackage{graphicx}
\usepackage[font=small,justification=centerlast,format=plain]{caption}
\usepackage{subcaption}
\usepackage{xcolor}
\usepackage{dsfont}
\usepackage{booktabs}
\usepackage{bm}
\usepackage[capitalise]{cleveref}
\usepackage{tikz}
\usepackage{pgf} 
\usepackage{color}
\definecolor{myblue}{rgb}{0.0, 0.0, 0.8}
\definecolor{myyellow}{rgb}{0.9, 0.9, 0.0}

\begin{document}

\title{Dynamics of Majority Rule  on Hypergraphs}
\author{James Noonan}
\email{jmt.noonan@gmail.com}
\affiliation{Mathematical Institute, University of Oxford, Oxford, UK}
\author{Renaud Lambiotte}
\email{renaud.lambiotte@maths.ox.ac.uk}
\affiliation{Mathematical Institute, University of Oxford, Oxford, UK}

\date{\today}

\begin{abstract}
A broad range of dynamical systems involve multi-body interactions, or group interactions, which may not be encoded in traditional graphical structures. In this work, we focus on a canonical example from opinion dynamics, the Majority Rule, and investigate the possibility to represent and analyse the system by means of hypergraphs. We explore the formation of consensus and restrict our attention to interaction groups of size $3$, in order to simplify our analysis from a combinatorial perspective. We propose different types of hypergraph  models, incorporating modular structure or degree heterogeneity, and 
recast the dynamics in terms of Fokker-Planck equations, which allows us to predict the transient dynamics toward consensus.  Numerical simulations  show a very good agreement between the stochastic dynamics and theoretical predictions for large population sizes.
\end{abstract}

\keywords{majority rule, consensus, higher-order, non-linear, networks, group dynamics, hypergraphs}

\maketitle

\section{Introduction}\label{Introduction}

Opinion dynamics is concerned with the study of consensus formation in populations of interacting individuals. An important evolution of the field has been to consider the impact of the structure of the underlying social network on the dynamics \cite{castellano2009statistical}. In contrast with mean-field approaches, where all the agents are essentially connected with each other, network-based approaches assume that the agents are located at the nodes of a network and that they are sparsely related by direct, binary connections through edges. A combination of these edges allows for indirect connections through the notion of path. A wide range of models have  been proposed so as to capture the ways in which social interactions between agents affect opinion formation, through factors like peer pressure and conviction. Here, we will focus on the popular class of models where each node may be in one of two states, represented by the binary variables $\textbf{0}$ and $\textbf{1}$. These allow us to model opinions relating to questions with a ``yes"/``no" answer, as in a referendum for instance, or ``left"/``right" political choices, and find interesting connections with statistical physics models for spin dynamics.

 The Voter Model (VM) is a prime example of such binary models  \cite{clifford1973model} and is defined as follows. At discrete times, an agent is chosen uniformly at random from the population. This agent then adopts the opinion of one of its randomly-chosen neighbours in the underlying network. This update is repeated \textit{ad infinitum}, or until consensus is necessarily reached on a finite connected graph. Importantly, VM is linear and it is dyadic in nature given that pairwise interactions alone are sufficient to capture the dynamics of the system. It is well known that VM is solvable on regular lattice structures in arbitrary spatial dimensions \cite{redner2001guide}. This is due to the fact that the average node state is conserved on degree-regular graphs. VM has also been shown to be conservative on heterogeneous networks \cite{sood2008voter}, allowing for significant analytical progress to be made in the study of the associated dynamics on a wide range of graph topologies. 

However, real-world dynamics often exhibit nonlinear and non-conservative behaviour. An important non-conservative generalisation of VM is the so-called Majority Rule (MR) model \cite{galam2002minority,krapivsky2003dynamics} where, at each update event, we choose a group of $G$ agents from the population, where $G\geq 3$. These agents form the interaction group, and may be chosen uniformly at random (as in a mean-field scenario), or in a way that is constrained by the underlying network, as we discuss below. 
After an interaction group is formed, all of its agents simultaneously adopt the majority opinion in the group. When $G$ is odd, the majority opinion is always well defined. When $G$ is even, and a tie is observed between the opinions, the consensus is either decided randomly, or by introducing a bias for one opinion \cite{friedman1985tyranny}. 

An important aspect of MR is that interactions take place in groups, motivated by the mechanism of peer pressure, which questions the adequacy of networks to encode the interactions between agents. Indeed the dynamical model is based on group interactions while the underlying network only encodes pairwise interactions, and there is thus 
no simple way to choose a group of $G$ agents, e.g. when $G=3$, should they form a triangle, or simply form a path of length 2? This type of questioning has gained a lot of attention in recent years \cite{lambiotte_2019,battiston2020networks}, as it was observed that many systems exhibit multi-body or  group interactions, such as in neural activity \cite{giusti_clique_2015,reimann_cliques_2017, santos_topological_2018}, robotics \cite{olfati2007consensus} or scientific collaborations \cite{patania_shape_2017}, and that traditional graphical structures are incapable of reflecting the multi-body nature of such interactions. In order to circumvent this issue, a natural choice is to adopt a more general class of topological structures known as \textit{hypergraphs} where interactions may, as the dynamical system, be multi-body. While there are other topological structures that are capable of encoding group interactions such as simplicial complexes \cite{torres2020simplicial} and bipartite graphs \cite{newman2002random}, hypergraphs constitute a straightforward representation of multi-body interactions on networks, and appear to be an ideal candidate for the representation and analysis of MR dynamics \cite{lanchier2013stochastic}. Importantly, as shown in  \cite{neuhauser2020opinion}, nonlinear multibody interactions are not expressible as linear combinations of pairwise interactions between adjacent nodes on a standard graph. MR dynamics on hypergraphs can thus not be reduced to a dynamics on standard networks in general. 

The main purpose of this article is to investigate MR dynamics on a broad range of hypergraph models. For the sake of simplicity, we will restrict our attention to interaction groups of size $3$, referred to as $triangles$, even if most of our results could be generalised to general group size. In Section \ref{MeanFieldAnalysis}, we start by reviewing results about the MR model in the mean-field limit, revealing its non-conservative nature and deriving some of its properties through a Fokker-Planck equation. In Section \ref{Tripartite}, we then investigate MR dynamics on the so-called tripartite hypergraph, which is a natural generalisation of the mean field case where the system is made of 3 types of nodes. In Section \ref{modularhypergraph}, we consider a model of hypergraphs with community structure, referred to as modular hypergraphs, and then extend our analysis to  heterogeneous hypergraphs in Section \ref{HeterogeneousNetworks}. Section \ref{Conclusion} concludes our work.
\section{Mean-Field Analysis}\label{MeanFieldAnalysis}
\subsection{Exact Analysis of the Exit Probability}
The selection rule constitutes a key element of the MR model. We begin by considering the dynamics in the mean-field. 
The model is defined as follows. Each node is endowed with a binary state variable, denoted by $\textbf{0}$ and $\textbf{1}$. At each time step, $3$ agents are chosen uniformly at random and the Majority Rule is applied. This random selection can be formulated conveniently in the language of hypergraphs. Let us consider a fully connected hypergraph of $N$ agents. The hypergraph structure $\mathcal{H}$  consists of the node set $V\left(\mathcal{H}\right)=\{1,\dots,N\}$ and, as we restrict ourselves to three-body interactions, the set of all possible triangles given by $T\left(\mathcal{H}\right)=\{\{i,j,k\}: i,j,k\in V\left(\mathcal{H}\right),i\neq j\neq k\}$. The resulting object is a first generalisation of fully connected graphs to hypergraphs. The selection of a group of 3 nodes is now defined as the random selection of one hyperedge in the set $T\left(\mathcal{H}\right)$ of available hyperedges.

When studying the Majority Rule, as well as other  opinion dynamics models with discrete states, an important quantity is the exit probability defined as follows.
 Suppose that the system is initialised with $n<N$ agents in the $\textbf{1}$ state and $N-n$ agents in the $\textbf{0}$ state. The exit probability $\mathcal{E}_{n}$ is the probability that the system reaches consensus with all agents in the $\textbf{1}$ state given that $n$ agents are initiated in the $\textbf{1}$ state. Krapivsky and Redner \cite{krapivsky2003dynamics} adopted a combinatorial approach in deriving an exact expression for $\mathcal{E}_{n}$ in the mean-field:
\begin{equation}\label{En}
    \mathcal{E}_{n} = \frac{1}{2^{N-3}}\sum_{j=1}^{n-1}\frac{\Gamma\left(N-2\right)}{\Gamma\left(j\right)\Gamma\left(N-j-1\right)}.
\end{equation}
%An explicit derivation of equation (\ref{En}) may be found in the Appendix.
\begin{figure}
\centering
    \includegraphics[width=0.8\linewidth]{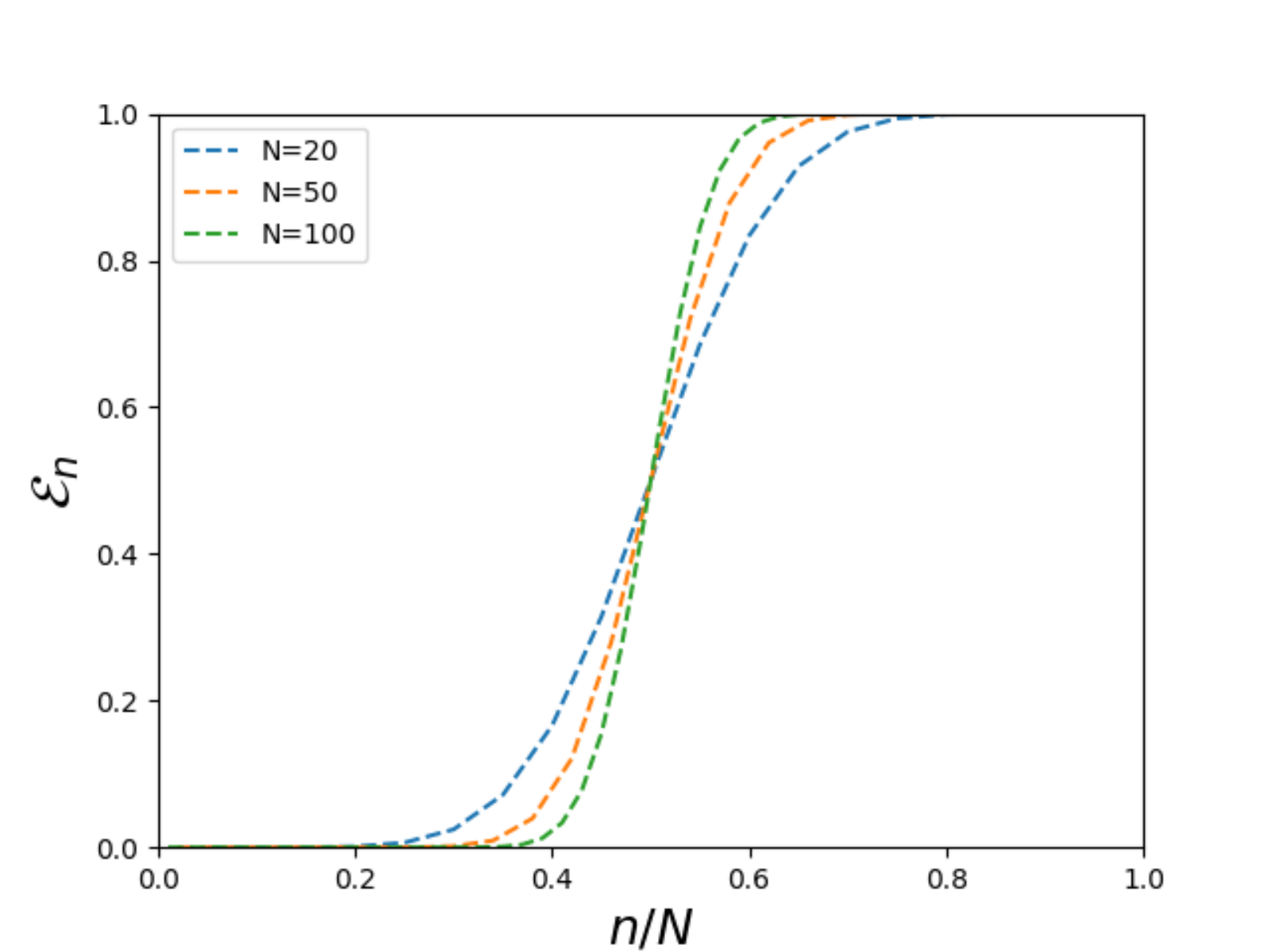}
    \caption[Mean field exit probability]{Exit probability for MR dynamics in the mean-field.}
    \label{ExitProbMeanField}
\end{figure}
Figure \ref{ExitProbMeanField} illustrates mean field exit probabilities using $N=20,50$ and $100$ agents. The initial fraction of agents $n/N$ in the $\textbf{1}$ state is plotted on the $x$-axis while the exit probability $\mathcal{E}_{n}$ is plotted on the $y$-axis. The exit probability is sigmoidal in nature, rapidly changing in a narrow interval centred at the pivotal value $n/N=0.5$. As $N\rightarrow \infty$ the exit probability converges to $0$ for $n/N<1/2$ \cite{krapivsky2003dynamics}. In other words, the asymptotic mean-field exit probability is akin to a step function with discontinuity at $n/N=0.5$. The complex functional form of $\mathcal{E}_{n}$ is a manifestation of the nonlinear and non-conservative nature of MR interactions.

 \subsection{The Fokker-Planck Approach}
 \noindent We now go beyond the workings presented in \cite{krapivsky2003dynamics} and approach the mean-field analysis from a different perspective. In this section, we study the exit probability and the dynamical approach towards consensus in the asymptotic limit  $N\rightarrow\infty$. Let $\rho\left(t\right)$ denote the density of \textbf{1}'s in the population at time $t$; that is, the fraction of agents in the $\textbf{1}$ state at time $t$. We shall write $\rho\left(t\right)$ as $\rho$ for convenience, though the dependence in $t$ is implicit. Furthermore, let $\delta\rho=N^{-1}$ denote the incremental change in $\rho$ following an update event. We  define the quantities $\rho^{\pm}=\rho\pm\delta\rho$ to reflect the density of states in the population after an interaction takes place. Let $R\left(\rho\right)$ and $L\left(\rho\right)$ denote the \textit{raising} and \textit{lowering operators} \cite{sood2008voter} that give the transition probabilities for the update events $\rho\rightarrow\rho^{+}$ and $\rho\rightarrow\rho^{-}$ respectively. Finally, let $p\left(\rho,t\right)$ be the probability that the density of $\textbf{1}$'s in the population is $\rho$ at time $t$. The probability density evolves according to the following master equation over the incremental time period $\delta t$:
 \begin{align}
     p\left(\rho,t+\delta t\right) &= R\left(\rho^{-}\right)p\left(\rho^{-},t\right)+L\left(\rho^{+}\right)p\left(\rho^{+},t\right) \nonumber \\ &+\left[1-R-L\right]p\left(\rho,t\right),\label{MFMasterEquation}
 \end{align}
 where $R$ and $L$ are assumed to denote $R\left(\rho\right)$ and $L\left(\rho\right)$ respectively. A conventional choice of the quantity $\delta t$ is $N^{-1}$ \cite{chen2005majority,sood2008voter}. We proceed by Taylor expanding the left hand side of equation (\ref{MFMasterEquation}) to first order in $\delta t$ and the right hand side to second order in $\delta p$. This ultimately yields the Fokker-Planck equation:
 \begin{align}
     \frac{\partial}{\partial t}p\left(\rho,t\right)&=-\frac{\delta\rho}{\delta t}\left[\frac{\partial}{\partial \rho}\left(R-L\right)p\left(\rho,t\right)\right] \nonumber \\&+\frac{\left(\delta\rho\right)^{2}}{2\delta t}\frac{\partial^{2}}{\partial\rho^{2}}\left[\left(R+L\right)p\left(\rho,t\right)\right].\label{MFFokkerPlanck}
 \end{align}
 
 In order to proceed, it is instructive to highlight the connection between equation (\ref{MFFokkerPlanck}) and its associated stochastic differential equation. Suppose that $X = \{X_{t}\left(\omega\right)\}_{t\in I,\omega \in \Omega}$ is a stochastic process on the sample space $\Omega$ over the time interval $I=\left[0,T\right]$ for some $T>0$, whose stochastic differential is given by
\begin{equation}\label{1DSDE}
    dX_{t}  = v\left(X_{t},t\right)dt + \sigma\left(X_{t},t\right)dW_{t}
\end{equation}
where $W_{t}$ is a standard Brownian motion. $v\left(X_{t},t\right)$ and $\sigma\left(X_{t},t\right)$ are referred to as the drift and diffusion coefficients respectively. We shall also refer to $v\left(X_{t},t\right)$ as the drift velocity. The Fokker-Planck equation associated with the probability density $p\left(x,t\right)$ of the random variable $X_{t}$ is given by
\begin{align}
    \frac{\partial }{\partial t}p\left(x,t\right) &= -\frac{\partial}{\partial x}\left[v\left(x,t\right)p\left(x,t\right)\right] \nonumber \\ &+ \frac{\partial^{2}}{\partial x^{2}}\left[D\left(x,t\right)p\left(x,t\right)\right]\label{1DFokkerPlanck}
\end{align}
where $D\left(X_{t},t\right) = \sigma^{2}\left(X_{t},t\right)/2$. Comparing equations (\ref{MFFokkerPlanck}) and (\ref{1DFokkerPlanck}) we observe that the density $\rho$ may be described as a stochastic process with drift velocity $v\left(\rho,t\right)$ and diffusion coefficient $D\left(\rho,t\right)$ defined as follows:
\begin{align}
    v\left(\rho,t\right)&=\frac{\delta\rho}{\delta t}\left(R-L\right)=R-L,\label{1dDrift} \\ D\left(\rho,t\right)&=\frac{\left(\delta\rho\right)^{2}}{2\delta t}\left(R+L\right)=\frac{1}{2N}\left(R+L\right).\label{1dDiffusion}
\end{align}
 If $N$ is sufficiently large so that $\left(N-1\right)/N\approx 1$, equations (\ref{1dDrift}) and (\ref{1dDiffusion}) simplify to give
\begin{align}
    v\left(\rho,t\right) &= 3\rho\left(1-\rho\right)\left(2\rho-1\right),\label{MFDrift}\\ D\left(\rho,t\right) &=\frac{3}{2N}\rho\left(1-\rho\right).\label{MFDiffusion}
\end{align}
Equations (\ref{MFDrift}) and (\ref{MFDiffusion}) illustrate that the drift velocity $v\left(\rho,t\right)$ is $O\left(1\right)$ whereas the diffusion term $D\left(\rho,t\right)$ is $O\left(1/N\right)$. The drift velocities thus dominate the dynamics of the system for sufficiently large $N$, as diffusive contributions vanish in the asymptotic limit. This observation will underpin much of the analysis presented throughout the course of this paper. 

\begin{figure}
     \centering
     \begin{subfigure}[b]{0.48\linewidth}
         \centering
         \includegraphics[width=\linewidth,height=\linewidth]{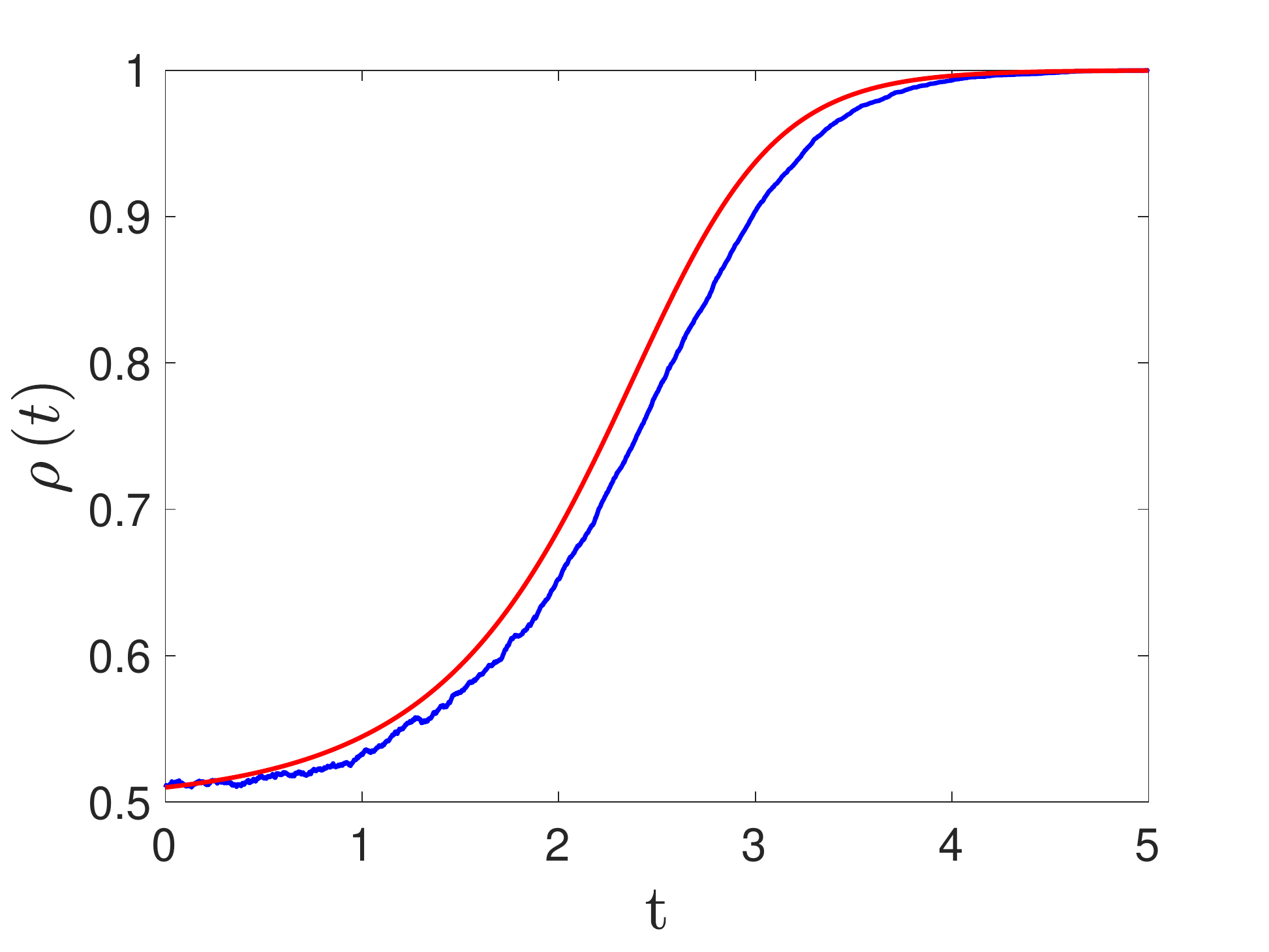}
         \caption{$\rho\left(0\right)=0.51$}
         \label{MFP51}
     \end{subfigure}
     \hfill
     \begin{subfigure}[b]{0.48\linewidth}
         \centering
         \includegraphics[width=\linewidth,height=\linewidth]{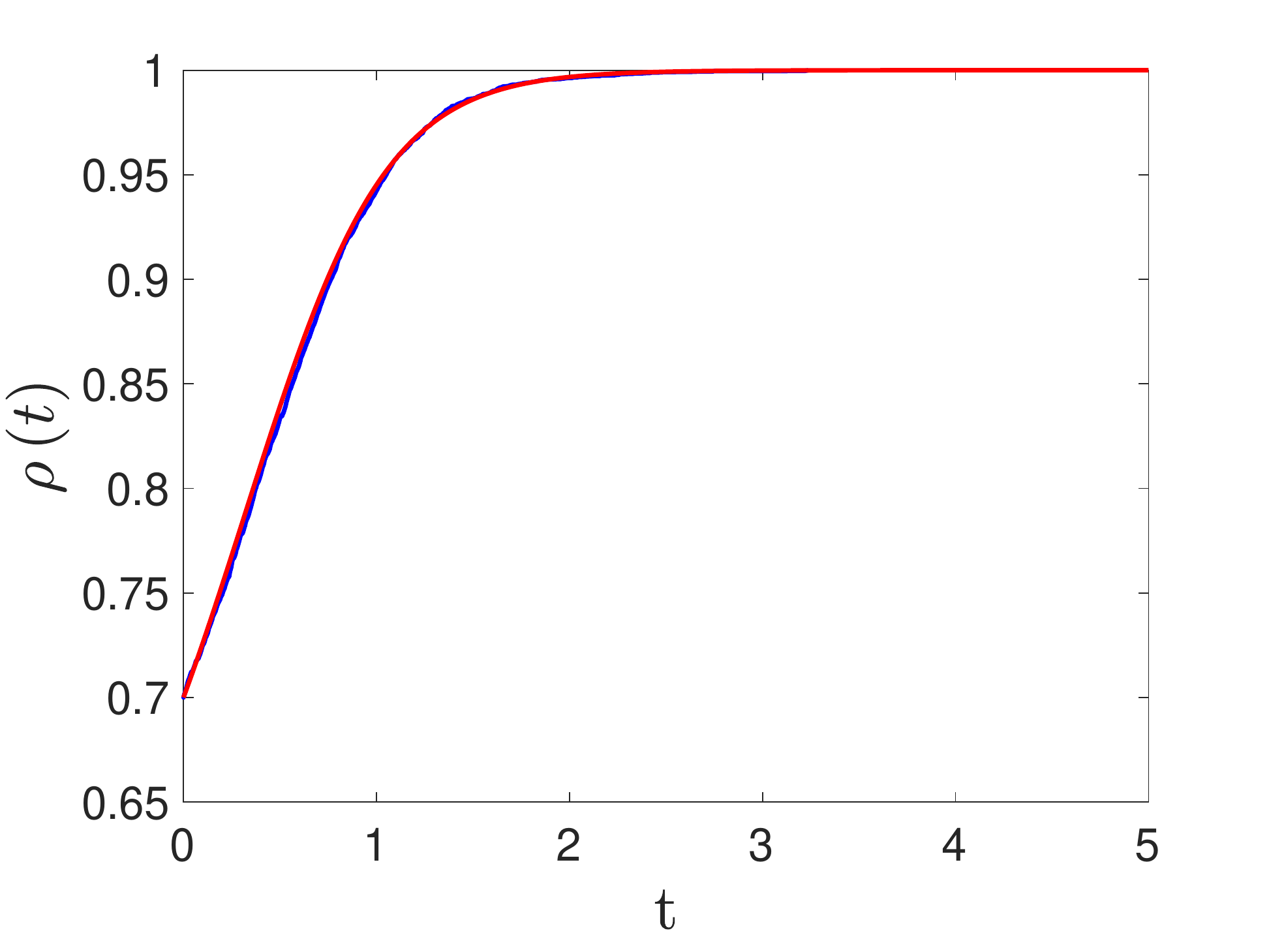}
         \caption{$\rho\left(0\right)=0.7$}
         \label{MFP7}
     \end{subfigure}
     \hfill
     %\begin{subfigure}[b]{0.45\linewidth}
     %    \centering
     %    \includegraphics[width=\linewidth,height=\linewidth]{TemporalMF9.eps}
     %    \caption{$\rho\left(0\right)=0.9$}
     %    \label{MFP9}
     %\end{subfigure}
        \caption[Mean-field temporal density profiles]{Temporal density profiles for mean-field MR dynamics with $N=10^{4}$.}
        \label{MRSimulations}
\end{figure}

 In the absence of diffusive contributions, we can use equation (\ref{1DSDE}) to write $v=d\rho/dt$, and thus integrate equation (\ref{MFDrift}) directly to give the following result:
\begin{equation*}\label{MFAnalytical}
    \rho\left(t\right)=\begin{cases}
    \frac{1}{2}\left(1-\sqrt{1-4/\left(4+\kappa e^{3t}\right)}\right)\quad\text{if}\quad \rho\left(0\right)<0.5, \\
    \frac{1}{2}\left(1+\sqrt{1-4/\left(4+\kappa e^{3t}\right)}\right)\quad\text{if}\quad \rho\left(0\right)>0.5,
    \end{cases}
\end{equation*}
where $\kappa=\left(2\rho\left(0\right)-1\right)^{2}/\left(\rho\left(0\right)\left(1-\rho\left(0\right)\right)\right)$. This indicates that the system will rapidly reach consensus with all agents in the $\textbf{0}$ state if $\rho\left(0\right)<0.5$. Conversely, if $\rho\left(0\right)>0.5$ then the system will reach consensus with all agents in the $\textbf{1}$ state. As a validation of our results, we simulate the mean-field MR dynamics in a population of size $N=10^{4}$. Figure \ref{MRSimulations} shows two representative results with $\rho\left(0\right)>0.5$. The density profiles are analogous for $\rho\left(0\right)<0.5$ with a reflection in the line $\rho=0.5$. The analytical trajectory is plotted in red, while the stochastic trajectory is plotted in blue. The time scale on the $x$-axis is measured in units of Monte Carlo steps per node, so that $\delta t=N^{-1}$. 
The trajectories are in excellent agreement when the initial value $\rho\left(0\right)$ is sufficiently far from $0.5$, and the system rapidly approaches consensus along the predicted trajectories.  When $\rho\left(0\right)=0.5\pm\epsilon$ where $0<\epsilon\ll 1$, however,  diffusive fluctuations at early times can lead to a temporal shift in the density profile along the $x$-axis,  and deviations between the predictions and the stochastic simulations can be observed, as illustrated in Figure \ref{MFP51}. Such diffusive effects become negligible as $N\rightarrow\infty$. 

%\noindent This concludes our mean field analysis of MR dynamics. To summarise, we followed the workings of Krapvisky and Redner \cite{krapivsky2003dynamics} in deriving an exact expression for the exit probability $\mathcal{E}_{n}$. We then provided some original workings by reformulating the analysis in terms of a Fokker-Planck equation, which allowed us to gain significant insight into the dynamics of the system for large $N$. We were able to verify our analysis by numerical means, showing that the temporal evolution of the stochastic dynamics is generally in excellent agreement with our predictions. We will now proceed to a discussion of MR dynamics on the tripartite hypergraph, which is a natural next step given its relatively simple topological structure.
\section{The Tripartite Hypergraph}\label{Tripartite}
In this Section we move beyond the mean-field  and consider hypergraphs with more sophisticated topologies. A natural starting point in this regard is the tripartite hypergraph, denoted by $\mathcal{H}$. The tripartite hypergraph consists of three distinct groups of nodes, which we shall refer to as $G_{a}$, $G_{b}$, and $G_{c}$. We assume for the sake of simplicity that each group consists of $N$ agents. Let $V_{a}\left(\mathcal{H}\right)=\{1_{a},\dots,N_{a}\},V_{b}\left(\mathcal{H}\right)=\{1_{b},\dots,N_{b}\},$ and $V_{c}\left(\mathcal{H}\right)=\{1_{c},\dots,N_{c}\}$ denote the sets of nodes in groups $G_{a},G_{b}$, and $G_{c}$ respectively. Let us define the set of triangles on $\mathcal{H}$ as $T\left(\mathcal{H}\right) = \{\{i,j,k\}:i\in V_{a}\left(\mathcal{H}\right),j\in V_{b}\left(\mathcal{H}\right),k\in V_{c}\left(\mathcal{H}\right)\}$. In other words, we consider all possible triplets with one agent in each group, hence generalising the notion of the complete bipartite graph. As before, at each step, a group of 3 nodes is chosen by selecting one hyperedge in $T\left(\mathcal{H}\right)$ at random.

Let us now adapt our treatment of the mean-field via a Fokker-Planck equation to this setting.
Let $\rho_{a}\left(t\right)$ denote the density of nodes in state $\textbf{1}$ in $G_{a}$ at time $t$, and let $\delta\rho_{a}=N^{-1}$ denote the change in $\rho_{a}\left(t\right)$ resulting from the changing of opinion of a single agent in $G_{a}$. We then define $\rho_{a}^{\pm}=\rho_{a}\pm\delta\rho_{a}$. Analogous expressions hold for agents in $G_{b}$ and $G_{c}$ where $\rho_{b}$ and $\rho_{c}$ denote the associated densities of $\textbf{1}$'s respectively. We also define the raising and lowering operators associated with $G_{a}$, denoted by $R_{a}\left(\rho_{a},\rho_{b},\rho_{c}\right)$ and $L_{a}\left(\rho_{a},\rho_{b},\rho_{c}\right)$ respectively (written as $R_{a}$ and $L_{a}$ for convenience). $R_{a}$ is the transition probability associated with the update event $\rho_{a}\rightarrow\rho_{a}^{+}$, whereas $L_{a}$ is the transition probability associated with the update event $\rho_{a}\rightarrow\rho_{a}^{-}$. Once again, analogous expressions exist for the raising and lowering operators associated with groups $G_{b}$ and $G_{c}$. $R_{a}$ and $L_{a}$ are given as follows:
\begin{align}
    R_{a} &= \left(1-\rho_{a}\right)\rho_{b}\rho_{c}, \label{RaTripartite} \\
    L_{a} &= \rho_{a}\left(1-\rho_{b}\right)\left(1-\rho_{c}\right). \label{LaTripartite}
\end{align}
Equation (\ref{RaTripartite}) follows from the fact that $\rho_{a}$ increases to $\rho_{a}^{+}$ if we choose a node in state $\textbf{0}$ from $G_{a}$ as well as two nodes in state $\textbf{1}$ from $G_{b}$ and $G_{c}$. Similarly, equation (\ref{LaTripartite}) is derived from the fact that $\rho_{a}$ decreases to $\rho_{a}^{-}$ if we choose a node in state $\textbf{1}$ from $G_{a}$ as well as two nodes in state $\textbf{0}$ from $G_{b}$ and $G_{c}$. The same logic applies for the calculation of the transition probabilities for groups $G_{b}$ and $G_{c}$. 

Using these transition probabilities we can deduce the probabilistic master equation governing MR dynamics on the tripartite hypergraph, where $p\left(\rho_{a},\rho_{b},\rho_{c},t\right)$ is the probability of the system having densities $\rho_{a},\rho_{b}$ and $\rho_{c}$ at time $t$:
\begin{align}
    p\left(t+\delta t\right) &= R_{a}\left(\rho_{a}^{-}\right)p\left(\rho_{a}^{-}\right)+L_{a}\left(\rho_{a}^{+}\right)p\left(\rho_{a}^{+}\right) \nonumber \\
    &+ R_{b}\left(\rho_{b}^{-}\right)p\left(\rho_{b}^{-}\right)+L_{a}\left(\rho_{b}^{+}\right)p\left(\rho_{b}^{+}\right) \nonumber \\
    &+ R_{c}\left(\rho_{c}^{-}\right)p\left(\rho_{c}^{-}\right)+L_{a}\left(\rho_{c}^{+}\right)p\left(\rho_{c}^{+}\right) \nonumber \\
    &+\left[1-R_{a}-R_{b}-R_{c}-L_{a}-L_{b}-L_{c}\right]p.\label{TripartiteMaster}
\end{align}
We proceed by Taylor expanding the left hand side of equation (\ref{TripartiteMaster}) to first order in $\delta t$ and the right hand side to second order in $\delta\rho_{a},\delta\rho_{b}$ and $\delta\rho_{c}$. We take $\delta t$ to be equal to the reciprocal of the total number of nodes in the hypergraph, in line with the convention introduced in Section \ref{MeanFieldAnalysis}. This ultimately yields the three-dimensional Fokker-Planck equation:
\begin{align}
    \frac{\partial p}{\partial t} \nonumber &=
     -\frac{\delta\rho_{a}}{\delta t}\frac{\partial}{\partial\rho_{a}}\left[\left(R_{a}-L_{a}\right)p\right]+\frac{\left(\delta\rho_{a}\right)^{2}}{2\delta t}\frac{\partial^{2}}{\partial\rho_{a}^{2}}\left[\left(R_{a}+L_{a}\right)p\right] \nonumber \\
    &-\frac{\delta\rho_{b}}{\delta t}\frac{\partial}{\partial\rho_{b}}\left[\left(R_{b}-L_{b}\right)p\right]+\frac{\left(\delta\rho_{b}\right)^{2}}{2\delta t}\frac{\partial^{2}}{\partial\rho_{b}^{2}}\left[\left(R_{b}+L_{b}\right)p\right] \nonumber \\
    &-\frac{\delta\rho_{c}}{\delta t}\frac{\partial}{\partial\rho_{c}}\left[\left(R_{c}-L_{c}\right)p\right]
    +\frac{\left(\delta\rho_{c}\right)^{2}}{2\delta t}\frac{\partial^{2}}{\partial\rho_{c}^{2}}\left[\left(R_{c}+L_{c}\right)p\right].\label{TripartiteFokkerPlanck}
\end{align}

As in the mean field case, it is instructive to highlight the connection between equation (\ref{TripartiteFokkerPlanck}) and its corresponding stochastic differential equation. Consider an $n$-dimensional stochastic process $\textbf{X}=\{\textbf{X}_{t}\left(\omega\right)\}_{t\in I,\omega\in\Omega}$ on the time interval $I=\left[0,T\right]$ and the sample space $\Omega$, where $n\geq1$ and $T>0$. Suppose $\textbf{X}_{t}\left(\omega\right)=\textbf{X}_{t}$ has stochastic differential
\begin{equation}\label{SDEmultidimensional}
    d\textbf{X}_{t}=\textbf{v}\left(\textbf{X}_{t},t\right)dt+\boldsymbol\sigma\left(\textbf{X}_{t},t\right)d\textbf{W}_{t},
\end{equation}
where $\textbf{v}\left(\textbf{X}_{t},t\right)$ is an $n$-dimensional random vector, $\boldsymbol\sigma\left(\textbf{X}_{t},t\right)$ is an $n\times m$ dimensional matrix and $\textbf{W}_{t}$ is a standard $m$-dimensional Weiner process where $m\geq 1$. If we denote the probability density associated with the stochastic process $\textbf{X}_{t}$ by $p\left(\textbf{x},t\right)$, then $p\left(\textbf{x},t\right)$ obeys the following Fokker-Planck equation:
\begin{align}
    \frac{\partial p\left(\textbf{x},t\right)}{\partial t}&=-\sum_{i=1}^{n}\frac{\partial}{\partial x_{i}}\left[v_{i}\left(\textbf{x},t\right)p\left(\textbf{x},t\right)\right] \nonumber\\ &+\sum_{i=1}^{n}\sum_{j=1}^{n}\frac{\partial^{2}}{\partial x_{i}\partial x_{j}}\left[D_{ij}\left(\textbf{x},t\right)p\left(\textbf{x},t\right)\right],\label{generalFokkerPlanck}
\end{align}
where $\textbf{v} = \left[v_{1},...,v_{n}\right]$ is the vector of drift velocities and $D=\frac{1}{2}\boldsymbol\sigma\boldsymbol\sigma^{T}$ is the diffusion tensor. Let us denote the drift velocity associated with the density of $\textbf{1}$'s in $G_{a}$ by $v_{a}$ and the corresponding diffusion term by $D_{a}$. 

\begin{figure*}
     \centering
     \begin{subfigure}[b]{0.3\linewidth}
         \centering
         \includegraphics[width=\linewidth,height=\linewidth]{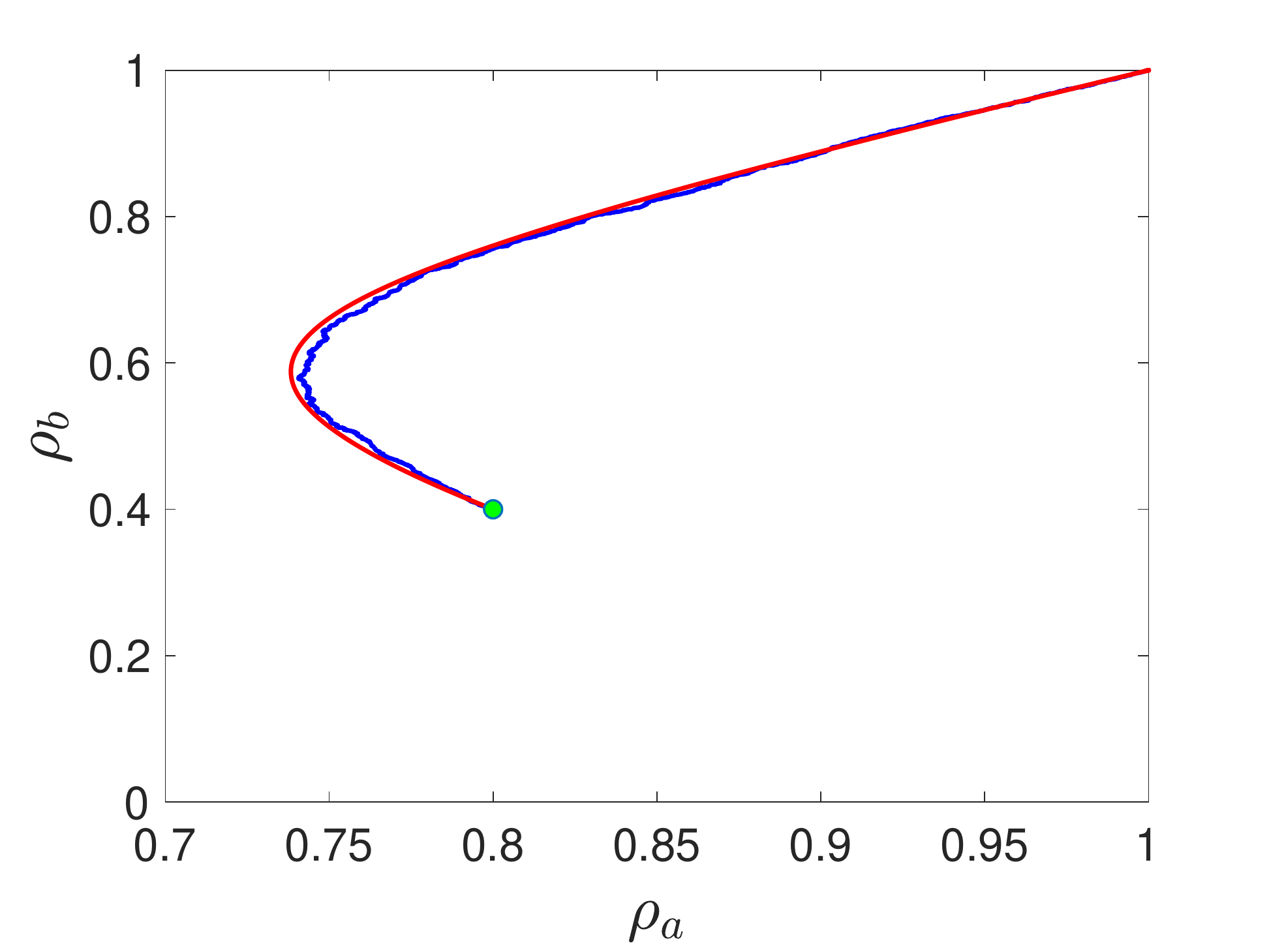}
         \caption{$\left(\rho_{a},\rho_{b}\right)$}
         \label{triPAPB}
     \end{subfigure}
     \hfill
     \begin{subfigure}[b]{0.3\linewidth}
         \centering
         \includegraphics[width=\linewidth,height=\linewidth]{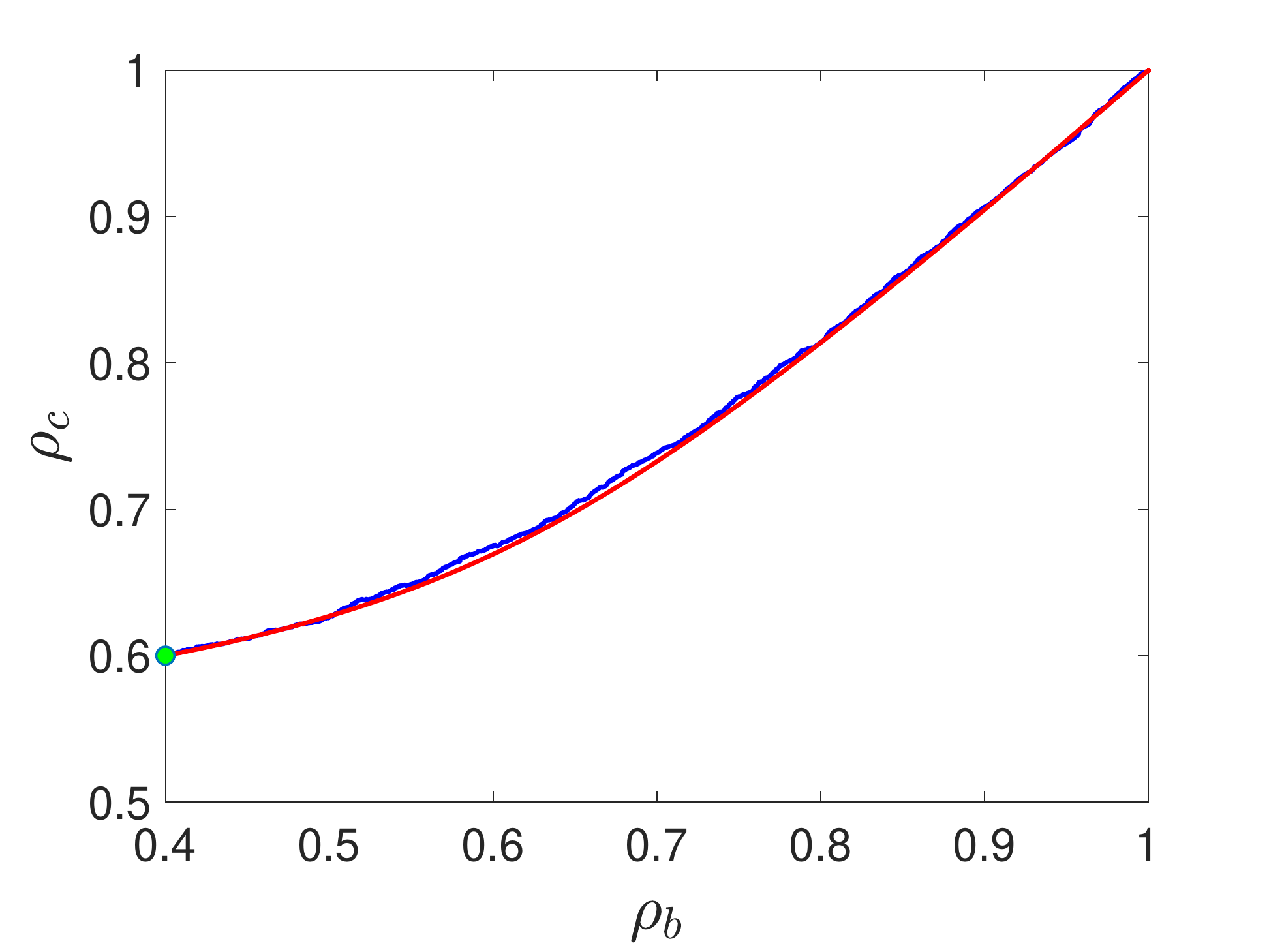}
         \caption{$\left(\rho_{b},\rho_{c}\right)$}
         \label{triPBPC}
     \end{subfigure}
     \hfill
     \begin{subfigure}[b]{0.3\linewidth}
         \centering
         \includegraphics[width=\linewidth,height=\linewidth]{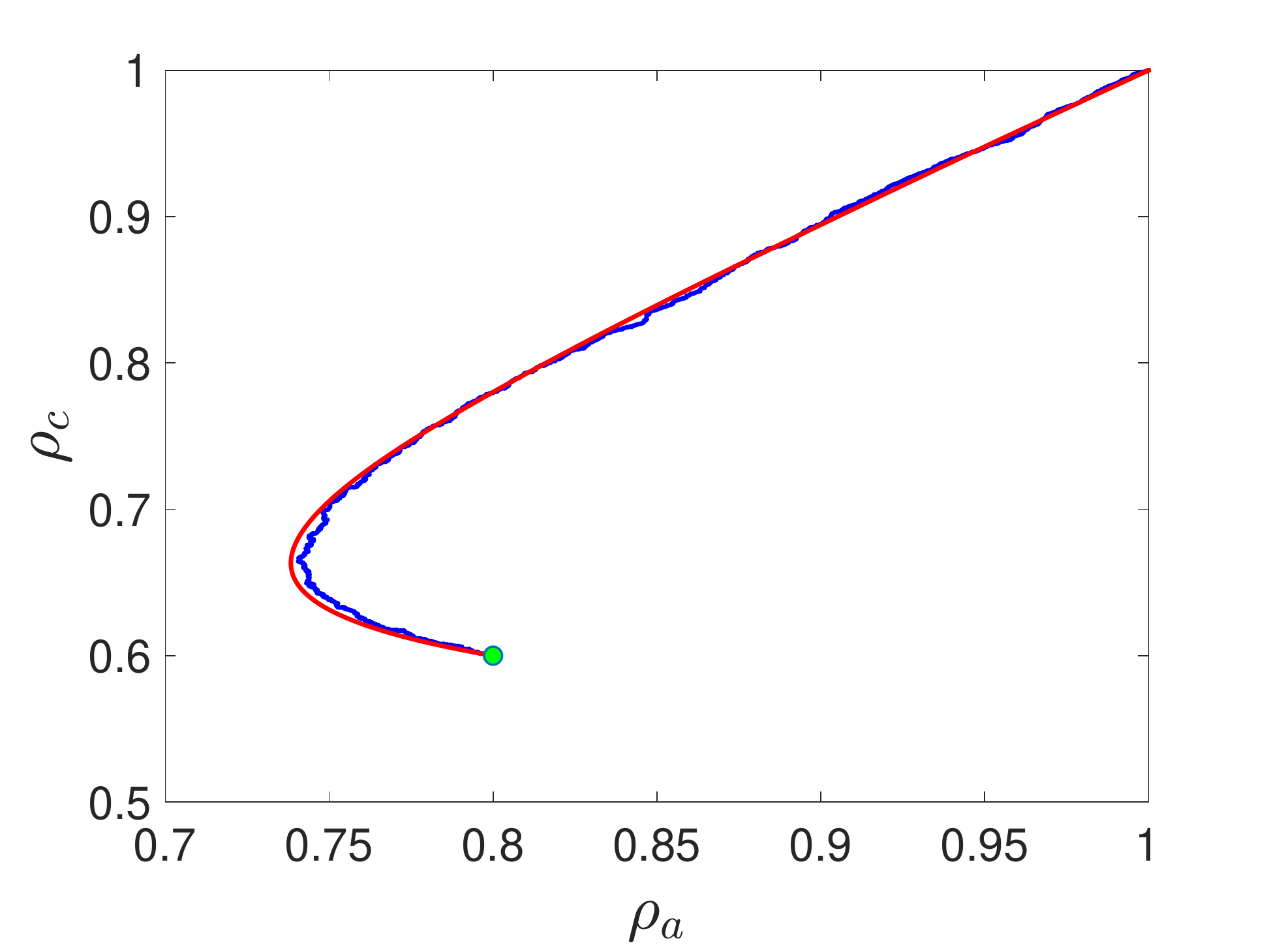}
         \caption{$\left(\rho_{a},\rho_{c}\right)$}
         \label{triPAPC}
     \end{subfigure}
        \caption[Simulations on the tripartite hypergraph]{Simulation results on the tripartite hypergraph with $N=10^{4}$. The initial conditions are given by $\boldsymbol\rho_{0}=\left(0.8,0.4,0.6\right)$, indicated by the green marker. Consensus is ultimately reached with all nodes in the \textbf{1} state.}
        \label{TripartiteProjections}
\end{figure*}

By comparing equations (\ref{TripartiteFokkerPlanck}) and (\ref{generalFokkerPlanck}) we can deduce the following expressions:
\begin{align}
    v_{a} &= 3\left[\left(1-\rho_{a}\right)\rho_{b}\rho_{c}-\rho_{a}\left(1-\rho_{b}\right)\left(1-\rho_{c}\right)\right], \label{TripartiteVa} \\
    D_{a} &= \frac{3}{2N}\left[\left(1-\rho_{a}\right)\rho_{b}\rho_{c}+\rho_{a}\left(1-\rho_{b}\right)\left(1-\rho_{c}\right)\right]. \label{TripartiteDa}
\end{align}
Note that in this instance the diffusion tensor is diagonal, where $D_{a}, D_{b}$, and $D_{c}$ are equal to $D_{11},D_{22}$, and $D_{33}$ respectively. This is because we are considering interaction groups of size $3$, meaning that only one of the quantities $\rho_{a},\rho_{b}$, or $\rho_{c}$ is varied at each update event. This is inherent in the structure of equation (\ref{TripartiteMaster}).  Analogous expressions exist for the drift and diffusion terms for $\rho_{b}$ and $\rho_{c}$. As in the mean field case, we note that the drift velocity is $O\left(1\right)$ whereas the diffusive term is $O\left(1/N\right)$. This implies that the dynamics are dominated by the drift velocities for large $N$, given that the diffusive contributions become negligible as $N\rightarrow\infty$. Using this fact, we posit that the dynamics of the system may be well approximated by discarding the diffusive terms and formulating a dynamical system using the drift velocities alone. This approach follows from the fact that in the absence of diffusive contributions, equation (\ref{SDEmultidimensional}) reduces to a system of ordinary differential equations:
\begin{align*}
    &\frac{d\rho_{a}}{dt} = v_{a} =  3\left[\left(1-\rho_{a}\right)\rho_{b}\rho_{c}-\rho_{a}\left(1-\rho_{b}\right)\left(1-\rho_{c}\right)\right], \\
    &\frac{d\rho_{b}}{dt} = v_{b} = 3\left[\left(1-\rho_{b}\right)\rho_{a}\rho_{c}-\rho_{b}\left(1-\rho_{a}\right)\left(1-\rho_{c}\right)\right],  \\
    &\frac{d\rho_{c}}{dt} = v_{c} = 3\left[\left(1-\rho_{c}\right)\rho_{a}\rho_{b}-\rho_{c}\left(1-\rho_{a}\right)\left(1-\rho_{b}\right)\right].
\end{align*}
This system may be analysed via linear stability analysis. The set of fixed points is given by $\{\left(0,0,0\right),\left(1/2,1/2,1/2\right),\left(1,1,1\right)\}$. When evaluated at the points $\left(0,0,0\right)$ and $\left(1,1,1\right)$ the eigenvalues of the Jacobian are $-3$ with multiplicity $3$, indicating asymptotic stability. When evaluated at $\left(1/2,1/2,1/2\right)$ the eigenvalues of the Jacobian are $-3$ with multiplicity $2$ and $3/2$ with multiplicity $1$. This implies that $\left(1/2,1/2,1/2\right)$ is a saddle point.

 We investigate the performance of our model by conducting simulations on a tripartite hypergraph of $N = 10^{4}$ nodes. Let us denote the initial conditions $\left(\rho_{a}\left(0\right),\rho_{b}\left(0\right),\rho_{c}\left(0\right)\right)$ by $\boldsymbol\rho_{0}$. Figure \ref{TripartiteProjections} illustrates a sample set of simulation results with $\boldsymbol\rho_{0}=\left(0.8,0.4,0.6\right)$. The trajectory predicted by our deterministic nonlinear system is plotted in red, whereas the stochastic trajectory resulting from simulating the MR dynamics is plotted in blue. Given that the dynamical system is three-dimensional, it is informative to study the projections of the resultant trajectories in the $\left(\rho_{a},\rho_{b}\right),\left(\rho_{b},\rho_{c}\right)$ and $\left(\rho_{a},\rho_{c}\right)$ planes. These projections are plotted from left to right in Figure \ref{TripartiteProjections}. The deterministic and stochastic trajectories are evidently in excellent agreement. This serves to validate our conjecture that the drift velocities dominate the dynamics of the system for large values of $N$.

\begin{figure*}
     \centering
     \begin{subfigure}[b]{0.3\linewidth}
         \centering
         \includegraphics[width=\linewidth,height=\linewidth]{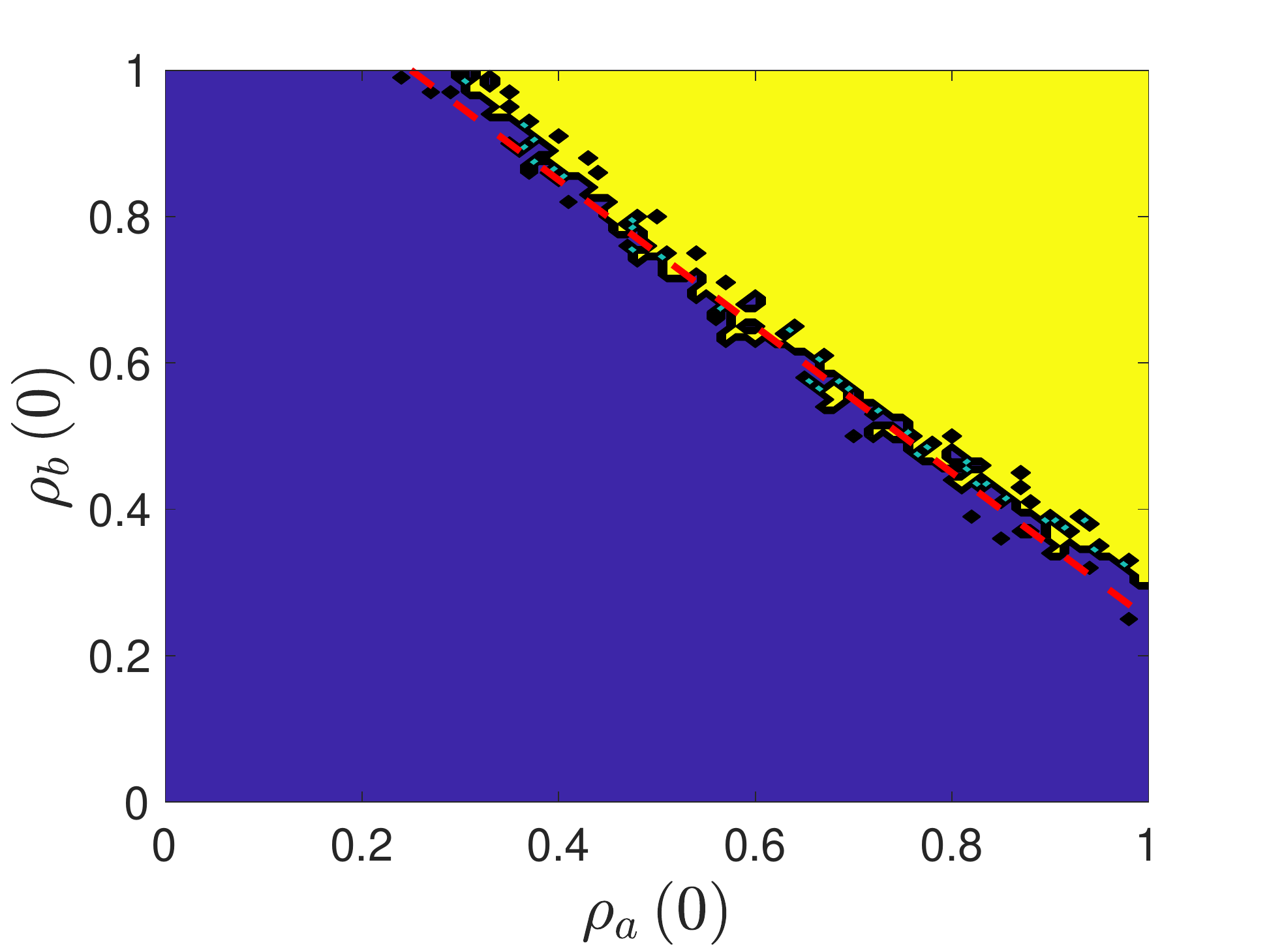}
         \caption{$\rho_{c}\left(0\right)=0.25$}
         \label{rhoc25}
     \end{subfigure}
     \hfill
     \begin{subfigure}[b]{0.3\linewidth}
         \centering
         \includegraphics[width=\linewidth,height=\linewidth]{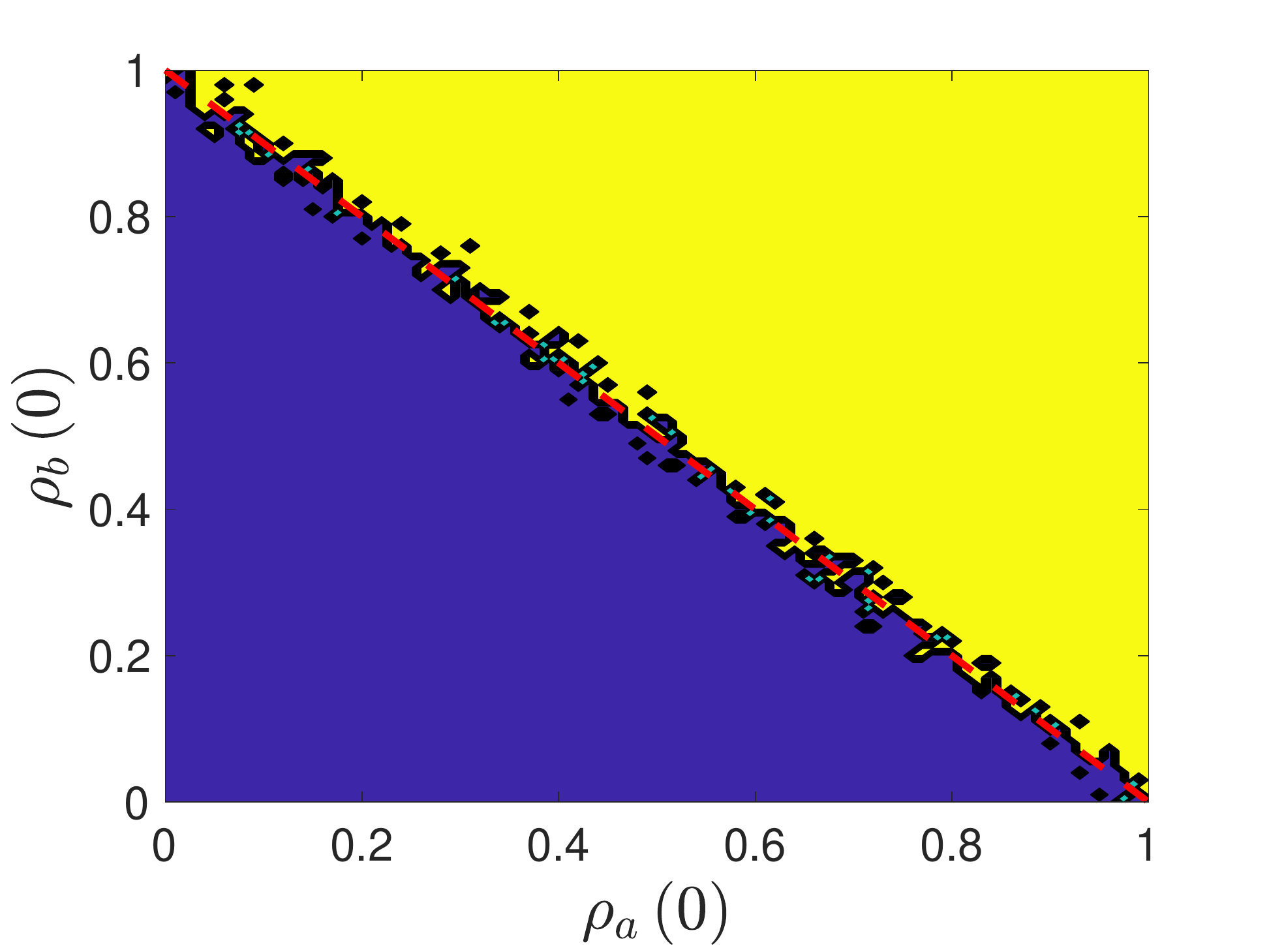}
         \caption{$\rho_{c}\left(0\right)=0.5$}
         \label{rhoc5}
     \end{subfigure}
     \hfill
     \begin{subfigure}[b]{0.3\linewidth}
         \centering
         \includegraphics[width=\linewidth,height=\linewidth]{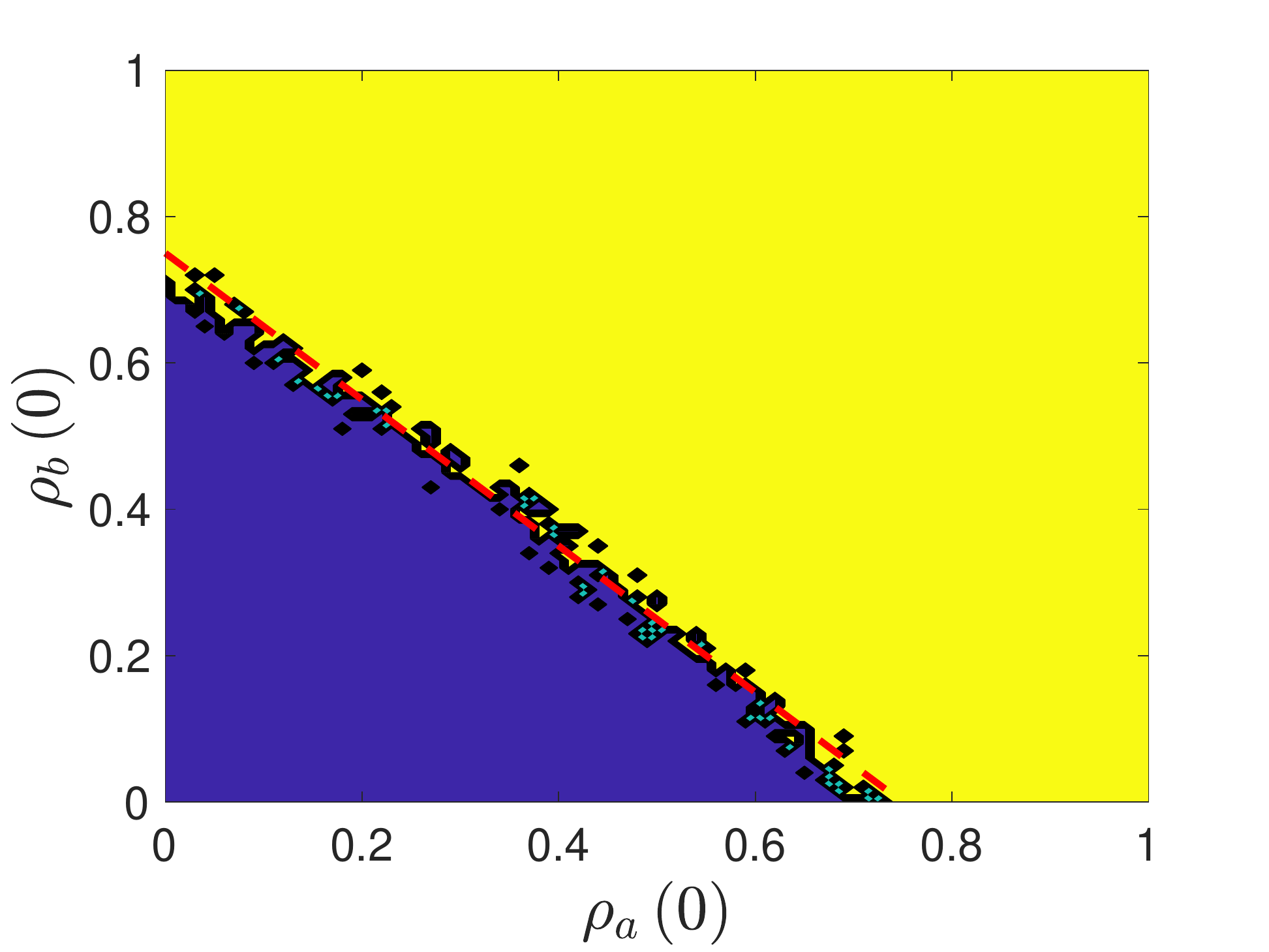}
         \caption{$\rho_{c}\left(0\right)=0.75$}
         \label{rhoc75}
     \end{subfigure}
        \caption[Exit probability analysis on the tripartite hypergraph]{Exit probability analysis on the tripartite hypergraph with $N=900$.}
        \label{TripartiteExitProb}
\end{figure*}

Our numerical simulations thus far have only focused on a specific trajectory of the dynamics resulting from an arbitrary choice of initial condition. We now extend our analysis by approximating the exit probability $\mathcal{E}\left(\boldsymbol\rho_{0}\right)$ of the system for large $N$; that is, the probability that all of the agents reach consensus in the \textbf{1} state if the initial conditions are given by $\boldsymbol\rho_{0}$. In order to proceed, we consider the unstable fixed point at $\left(1/2,1/2,1/2\right)$. The eigenvectors of the Jacobian when evaluated at this point are mutually orthogonal. The unstable linear subspace $E^{u}$ is spanned by the vector $\left[1,1,1\right]$, which is normal to the surface of the stable linear subspace $E^{s}$ spanned by the remaining two eigenvectors. It is straightforward to deduce that the equation of the plane corresponding to the stable linear subspace is given by
\begin{equation}\label{Estable}
    \rho_{a}+\rho_{b}+\rho_{c} = 3/2.
\end{equation}
By symmetry we posit that in the limit of large $N$, the exit probability is given by the following piecewise function:
\begin{equation}
    \mathcal{E}\left(\boldsymbol\rho_{0}\right)=
    \begin{cases}
    0\quad\text{if}\quad \rho_{a}\left(0\right)+\rho_{b}\left(0\right)+\rho_{c}\left(0\right)<3/2, \\
    1\quad\text{if}\quad \rho_{a}\left(0\right)+\rho_{b}\left(0\right)+\rho_{c}\left(0\right)>3/2.
    \end{cases}
\end{equation}
In order to confirm this prevision numerically, we vary $\rho_{a}\left(0\right)$ and $\rho_{b}\left(0\right)$ while keeping $\rho_{c}\left(0\right)$ fixed. Figure \ref{TripartiteExitProb} gives the simulation results for $\rho_{c}\left(0\right)=0.25,0.5,$ and $0.75$, where $N=900$. These plots were generated by initialising the stochastic dynamics at each point on the $\left(\rho_{a}\left(0\right),\rho_{b}\left(0\right)\right)$ grid with uniform mesh size $0.01$. Purple indicates consensus in the $\textbf{0}$ state whereas yellow indicates consensus in the $\textbf{1}$ state. The red dashed line marks the boundary predicted by equation (\ref{Estable}) at $t=0$. Only one simulation of the stochastic dynamics was conducted per coordinate, giving rise to diffusive effects at the interface separating the two domains. Nevertheless, the deterministic exit probability is in excellent agreement with the stochastic simulations of the system. 

%\noindent This concludes our analysis of MR dynamics on the tripartite hypergraph. The drift velocities proved to be an excellent predictor of the dynamical behaviour of the system for large $N$. The densities $\rho_{a},\rho_{b}$ and $\rho_{c}$ were observed to follow hyperbolic trajectories in phase space, allowing us to predict not only the ultimate consensus state of the system but also the way in which the system reaches consensus. In order to extend this analysis to more complicated hypergraph topologies, we proceed to a discussion of MR dynamics on hypergraphs with community structure in the next Chapter. 
\section{The Modular Hypergraph}\label{modularhypergraph}
\subsection{The Symmetric Case}\label{SymmetricModularText}
In this Section we consider MR dynamics on hypergraphs with community structure, otherwise known as modular hypergraphs \cite{kumar2018hypergraph,chodrow2020annotated}. In our setting, the population is partitioned into a number of non-overlapping sets known as modules or communities \cite{newman2006modularity,blondel2008fast}. The modular hypergraph $\mathcal{H}$ consists of two sets of $N$ nodes, referred to as communities $A$ and $B$ respectively. Let $V_{A}\left(\mathcal{H}\right)=\{1_{A},\dots,N_{A}\}$ and $V_{B}\left(\mathcal{H}\right)=\{1_{B},\dots,N_{B}\}$ denote the node sets in communities $A$ and $B$ respectively. The set of triangles on $\mathcal{H}$ is defined as $T\left(\mathcal{H}\right)=\{\{i,j,k\}:i,j,k\in V_{A}\left(\mathcal{H}\right)\cup V_{B}\left(\mathcal{H}\right),i\neq j\neq k\}$. In order to distribute the triangles over the hypergraph, we define the parameter $p_{ab}$ which gives the probability of forming a triangle with $a$ agents in A and $b$ agents in $B$. It follows that $p_{ab}\in\{p_{30},p_{21},p_{12},p_{03}\}$. These can be interpreted as hyperedge selection probabilities. Here, the hypergraph is assumed to be symmetric, meaning $p_{30}=p_{03}$ and $p_{21}=p_{12}$. 
The total number of triangles $T = |T\left(\mathcal{H}\right)|$ is given by 
\begin{equation*}
    T = \frac{2N\left(2N-1\right)\left(2N-2\right)}{3!} \approx \frac{4N^{3}}{3}.
\end{equation*}

 In order to vary the degree to which the two components of the modular hypergraph interact, we introduce a parametric dependence inspired by the study of VM dynamics on the two-clique graph \cite{sood2008voter}.
In graphical terms, communities $A$ and $B$ may be thought of as two fully-connected cliques of $N$ nodes. We then define an interconnectivity parameter $C\in\left[0,N\right]$ such that each node in clique $A$ is, on average, connected to $C$ nodes in clique $B$ and vice versa.  It follows that the average degree of a node in the graphical representation is $N+C$, as each clique is complete. Using this graphical representation it is straightforward to calculate hyperedge selection probabilities:
\begin{align}
    p_{30}&=p_{03}=\frac{1}{2}\left(\frac{N}{N+C}\right)^{2},\label{selectionProbabilities1} \\
    p_{21}&=p_{12}=\frac{NC}{\left(N+C\right)^{2}}+\frac{1}{2}\left(\frac{C}{N+C}\right)^{2}.\label{selectionProbabilities2}
\end{align}
These probabilities satisfy normalisation. When $C=0$ it follows that $p_{30}=p_{03}=1/2$ while $p_{12}=p_{21}=0$, hence the two modules evolve independently. Conversely, when $C=N$ it follows that $p_{30}=p_{03}=1/8$ whereas $p_{12}=p_{21}=3/8$, corresponding to a mean-field of $2N$ agents. We now define $N_{ab}$ as the number of hyperedges with $a$ nodes in $A$ and $b$ nodes in $B$. Using the fact that $p_{ab} = N_{ab}/T$ the hyperedge distribution may be determined as a function of $C$:
\begin{align}
    &N_{30} = N_{03} = \frac{2N^{5}}{3\left(N+C\right)^{2}}, \label{symHyperSel1} \\
    &N_{12} = N_{21} = \frac{4N^{3}C}{3\left(N+C\right)^{2}}\left(N+\frac{C}{2}\right).\label{symHyperSel2}
\end{align}
Note that hyperedges may have multiplicity greater than $1$. This is due to the fact that the total number of hyperedges $T$ is fixed, regardless of the value of $C$. For example, when $C=0$ we find that $N_{30}=N_{03}=2N^{3}/3$ whereas $N_{12}=N_{21}=0$. However, there are only $N^{3}/6$ ways to choose a hyperedge of size $3$ from $N$ nodes, implying that all hyperedges have multiplicity $4$ in this instance. This combinatorial consequence is immaterial to the dynamics of the system.
%Substituting $C=0$ gives $p_{30}=p_{03}=1/2$ and $p_{12}=p_{21}=0$, illustrating that the two communities evolve independently until they reach local consensus. In contrast, when $C=N$ we find $p_{30}=p_{03}=1/8$ and $p_{12}=p_{21}=3/8$ which corresponds to a fully-connected hypergraph in which each node belongs to one of the two cliques $A$ and $B$.

% The system evolves as follows at each update event:
%\begin{enumerate}
%    \item Choose an agent uniformly at random from the population. We shall refer to the clique in which the first agent resides as the \textit{primary} clique. The other clique is referred to as the \textit{secondary} clique.
%    \item Choose a further two agents to form the triangle. The probability of choosing two agents from the primary clique is $N^{2}/\left(N+C\right)^{2}$. The probability of choosing two agents from the secondary clique is $C^{2}/\left(N+C\right)^{2}$. The probability of choosing the second and third agents to be in different cliques is $2CN/\left(N+C\right)^{2}$ where the factor of $2$ accounts for different orderings.
%    \item All agents in the triangle adopt the majority opinion. 
%\end{enumerate}
%This process is repeated until the system reaches consensus. Note that triangle formation probabilities are calculated assuming that agents are chosen with replacement from the population. This is justified by the fact that for sufficiently large $N$, the probability of choosing the same agent twice is small.  
Once again, each agent in the population is assumed to occupy one of two states, labelled \textbf{0} and \textbf{1}. Simulations are conducted using the hyperedge selection probabilities given in equations (\ref{selectionProbabilities1}) and (\ref{selectionProbabilities2}). This is equivalent to choosing hyperedges uniformly at random from the hyperdegree distribution given by equations (\ref{symHyperSel1}) and (\ref{symHyperSel2}). Let us denote the density of nodes in state $\textbf{1}$ in $A$ and $B$ at time $t$ by $\rho_{A}\left(t\right)$ and $\rho_{B}\left(t\right)$ respectively. We write $\rho_{A}\left(t\right)$ as $\rho_{A}$ and $\rho_{B}\left(t\right)$ as $\rho_{B}$ for convenience. Let $\rho_{A}^{\pm} = \rho_{A}\pm \delta\rho_{A}$ and $\rho_{B}^{\pm} = \rho_{B}\pm\delta\rho_{B}$ where $\delta\rho_{A}=N^{-1}$ and $\delta\rho_{B}=N^{-1}$. Let $R_{A}\left(\rho_{A},\rho_{B}\right)$ and $R_{B}\left(\rho_{A},\rho_{B}\right)$ be the raising operators that give the transition probabilities from $\rho_{A}$ and $\rho_{B}$ to $\rho_{A}^{+}$ and $\rho_{B}^{+}$ respectively. Similarly, let $L_{A}\left(\rho_{A},\rho_{B}\right)$ and $L_{B}\left(\rho_{A},\rho_{B}\right)$ be the lowering operators that give the transition probabilities from $\rho_{A}$ and $\rho_{B}$ to $\rho_{A}^{-}$ and $\rho_{B}^{-}$ respectively. 
%We shall write $R_{A}\left(\rho_{A},\rho_{B}\right)$ as $R_{A}$, and similarly for the other operators. 
%For $R_{A}$, we note that there are three configurations of nodes that will give rise to an increase in $\rho_{A}$: $\left(\textbf{1}_{A},\textbf{1}_{A},\textbf{0}_{A}\right), \left(\textbf{1}_{A},\textbf{1}_{B},\textbf{0}_{A}\right)$, and $\left(\textbf{0}_{A},\textbf{1}_{B},\textbf{1}_{B}\right)$, where $\textbf{0}_{A}$ refers to a node in state $\textbf{0}$ in $A$, etc. Note that triangle formation probabilities generally depends on which node is chosen first. For example, consider the triplet $\left(\textbf{1}_{A},\textbf{1}_{B},\textbf{0}_{A}\right)$. If we choose $\textbf{1}_{A}$ as the first node then the probability of choosing another node from $A$ and a node from $B$ to form the triangle is given by $2NC/\left(N+C\right)^{2}$. However, if we choose $\textbf{1}_{B}$ as the  first node, then the probability of choosing two nodes from $A$ to form the triangle is given by $C^{2}/\left(N+C\right)^{2}$. 
$R_{A}$ and $L_{A}$ are given by
\begin{align}
    &R_{A} = \frac{3}{2}\left(\frac{N}{N+C}\right)^{2}\rho_{A}^{2}\left(1-\rho_{A}\right) \nonumber\\ &+ \frac{C\left(2N+C\right)}{\left(N+C\right)^{2}}\rho_{A}\rho_{B}\left(1-\rho_{A}\right) \nonumber\\
    &+ \frac{C\left(2N+C\right)}{2\left(N+C\right)^{2}}\rho_{B}^{2}\left(1-\rho_{A}\right), \label{RaSymmetric}\\
    &L_{A} = \frac{3}{2}\left(\frac{N}{N+C}\right)^{2}\rho_{A}\left(1-\rho_{A}\right)^{2} \nonumber\\ &+\frac{C\left(2N+C\right)}{\left(N+C\right)^{2}}\rho_{A}\left(1-\rho_{A}\right)\left(1-\rho_{B}\right) \nonumber \\&+\frac{C\left(2N+C\right)}{2\left(N+C\right)^{2}}\rho_{A}\left(1-\rho_{B}\right)^{2}. \label{RbSymmetric}
\end{align}

Analogous expression exist for the operators $R_{B}$ and $L_{B}$. Let $p\left(\rho_{A},\rho_{B},t\right)$ be the probability that communities $A$ and $B$ have densities $\rho_{A}$ and $\rho_{B}$ respectively at time $t$. Its associated master equation is given by
\begin{align}\label{masterprobSymmetric}
    p\left(t+\delta t\right) &=  R_{A}\left(\rho_{A}^{-}\right)p\left(\rho_{A}^{-}\right) + L_{A}\left(\rho_{A}^{+}\right)p\left(\rho_{A}^{+}\right) \nonumber \\
    &+ R_{B}\left(\rho_{B}^{-}\right)p\left(\rho_{B}^{-}\right) + L_{B}\left(\rho_{B}^{+}\right)p\left(\rho_{B}^{+}\right) \nonumber \\
    &+\left[1-R_{A}-L_{A}-R_{B}-L_{B}\right]p,
\end{align}
where $\delta t = \left(2N\right)^{-1}$.
 Taylor expanding the left hand side of equation (\ref{masterprobSymmetric}) to first order in $\delta t$ and the right hand side to second order in $\delta\rho_{A}$ and $\delta\rho_{B}$ yields the following Fokker-Planck equation:
\begin{align}\label{FokkerPlanckSymmetric}
    \frac{\partial p}{\partial t} &= -\frac{\delta\rho_{A}}{\delta t}\frac{\partial}{\partial \rho_{A}}\left[\left(R_{A}-L_{A}\right)p\right]-\frac{\delta\rho_{B}}{\delta t}\frac{\partial}{\partial \rho_{B}}\left[\left(R_{B}-L_{B}\right)p\right] \nonumber \\
    &+ \frac{\left(\delta\rho_{A}\right)^{2}}{2\delta t}\frac{\partial^{2}}{\partial \rho_{A}^{2}}\left[\left(R_{A}+L_{A}\right)p\right] \nonumber\\ &+\frac{\left(\delta\rho_{B}\right)^{2}}{2\delta t}\frac{\partial^{2}}{\partial \rho_{B}^{2}}\left[\left(R_{B}+L_{B}\right)p\right].
\end{align}

\begin{figure}
\centering
    \includegraphics[width=0.9\linewidth]{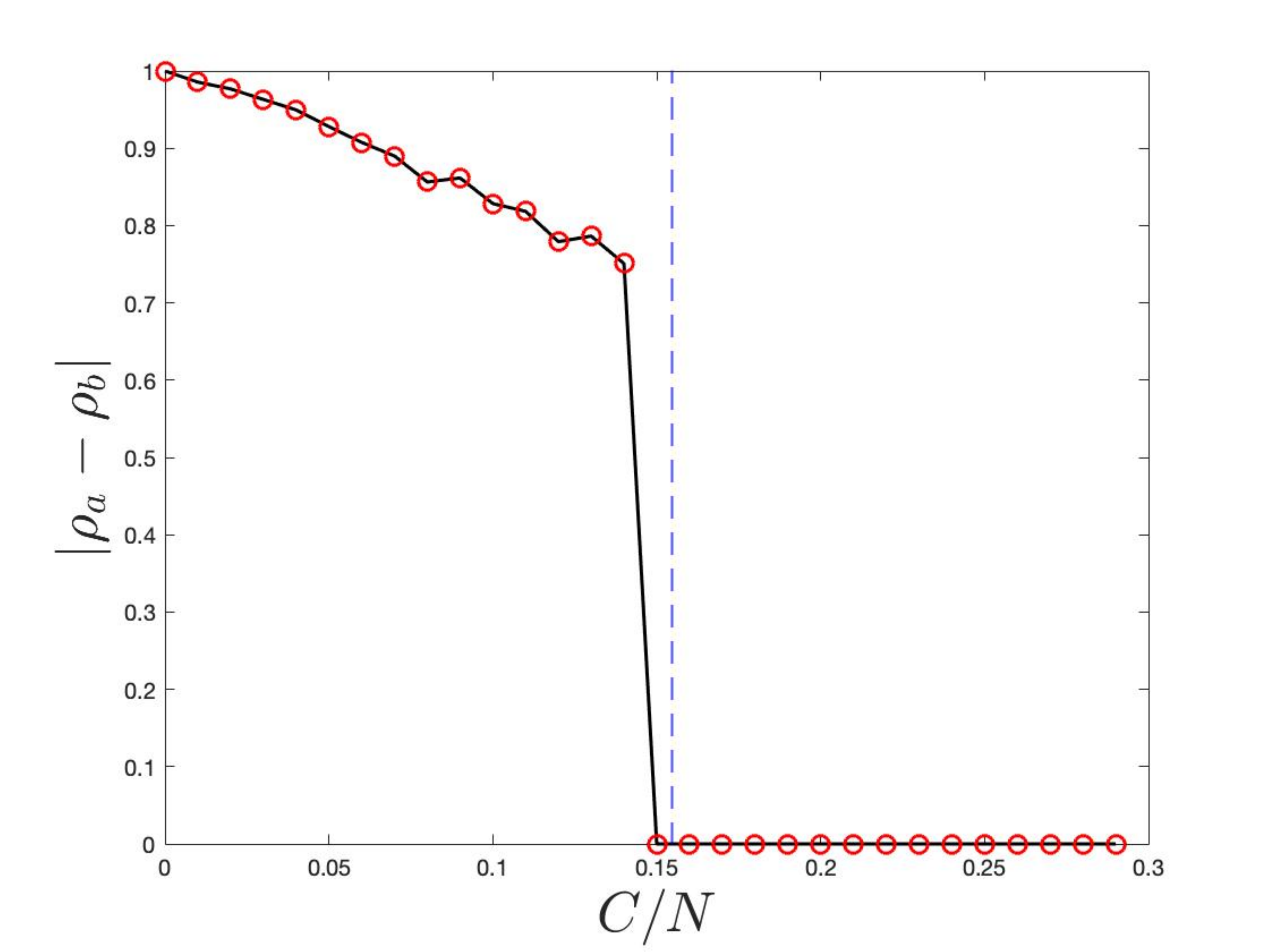}
    \caption[]{Phase diagram for the symmetric modular hypergraph with $N=2500$. The initial densities of $\textbf{1}'s$ in $A$ and $B$ were taken to be $1$ and $0$ respectively, with a different consensus in each community. The dashed line indicates the theoretical transition value $C^{t}$. Data points were averaged over $50$ simulations and $500N$ update events were conducted per simulation to ensure distributional stationarity. Simulations confirm a transition between the coexistence of different opinions  in different communities to global consensus when  $C$ is increased. Note that global consensus is, for any value of $C$, an absorbing state.}
    \label{PhaseDiagram}
\end{figure}

The drift velocities $v_{A}$ and $v_{B}$ of the clique densities $\rho_{A}$ and $\rho_{B}$ may be identified using equation (\ref{FokkerPlanckSymmetric}):
\begin{align}
    v_{A} &= 3\left(\frac{N}{N+C}\right)^{2}\rho_{A}\left(1-\rho_{A}\right)\left(2\rho_{A}-1\right) \nonumber \\ &+\frac{2C\left(2N+C\right)}{\left(N+C\right)^{2}}\rho_{A}\left(1-\rho_{A}\right)\left(2\rho_{B}-1\right) \nonumber \\&+\frac{C\left(2N+C\right)}{\left(N+C\right)^{2}}\left(\rho_{B}^{2}\left(1-\rho_{A}\right)-\rho_{A}\left(1-\rho_{B}\right)^{2}\right), \label{vaSymmetric}\\
    v_{B} &= 3\left(\frac{N}{N+C}\right)^{2}\rho_{B}\left(1-\rho_{B}\right)\left(2\rho_{B}-1\right) \nonumber \\
    &+\frac{2C\left(2N+C\right)}{\left(N+C\right)^{2}}\rho_{B}\left(1-\rho_{B}\right)\left(2\rho_{A}-1\right) \nonumber\\ &+\frac{C\left(2N+C\right)}{\left(N+C\right)^{2}}\left(\rho_{A}^{2}\left(1-\rho_{B}\right)-\rho_{B}\left(1-\rho_{A}\right)^{2}\right). \label{vbSymmetric}
\end{align}
Equations (\ref{vaSymmetric}) and (\ref{vbSymmetric}) imply that the drift velocities are $O\left(1\right)$ for all $C$. However, as $C$ tends to $0$ the equations decouple, implying that the two communities evolve almost independently of one another. When $C=0$ the drift velocities decouple completely. Using equation (\ref{FokkerPlanckSymmetric}) it is straightforward to show that the corresponding diffusion terms $D_{A}$ and $D_{B}$ are $O\left(1/N\right)$ for all $C$.
%\begin{align}
%    D_{A} &= \frac{\left(\delta\rho_{A}\right)^{2}}{2\delta t}\left(R_{A}+L_{A}\right) = \frac{3}{2}\frac{N}{\left(N+C\right)^{2}}\rho_{A}\left(1-\rho_{A}\right) \nonumber \\ 
%    &+\frac{C\left(2N+C\right)}{N\left(N+C\right)^{2}}\left[\rho_{A}\left(1-\rho_{A}\right)+\frac{1}{2}\left(\rho_{B}^{2}\left(1-\rho_{A}\right)+\rho_{A}\left(1-\rho_{B}\right)^{2}\right)\right], \label{DaSymmetric} \\
%    D_{B} &= \frac{\left(\delta\rho_{B}\right)^{2}}{2\delta t}\left(R_{B}+L_{B}\right) = \frac{3}{2}\frac{N}{\left(N+C\right)^{2}}\rho_{B}\left(1-\rho_{B}\right) \nonumber\\ 
%    &+\frac{C\left(2N+C\right)}{N\left(N+C\right)^{2}}\left[\rho_{B}\left(1-\rho_{B}\right)+\frac{1}{2}\left(\rho_{A}^{2}\left(1-\rho_{B}\right)+\rho_{B}\left(1-\rho_{A}\right)^{2}\right)\right]. \label{DbSymmetric}
%\end{align}
%Equations (\ref{DaSymmetric}) and (\ref{DbSymmetric}) illustrate that the diffusion terms are $O\left(1/N\right)$ for all $C$. As $C$ goes to $0$ we observe that equations (\ref{DaSymmetric}) and (\ref{DbSymmetric}) begin to decouple. When $C=0$ the diffusion terms decouple completely, as expected. \newline
This implies that the drift velocities dominate the dynamics of the system for relatively high values of $C$, allowing for diffusive contributions to be discarded. We approximate the stochastic dynamics using a deterministic model by writing $d\rho_{A}/dt=v_{A}$ and $d\rho_{B}/dt=v_{B}$ where $v_{A}$ and $v_{B}$ are given in equations (\ref{vaSymmetric}) and (\ref{vbSymmetric}) respectively. The associated linear stability analysis may be found in Appendix \ref{modularAppendix}. The fixed points at $\left(0,0\right)$ and $\left(1,1\right)$ are found to be stable for all $C$. On the other hand, $\left(1/2,1/2\right)$ is a saddle point for $C>C^{*}$ and a source for $C<C^{*}$ where $C^{*}\approx 0.2649N$. When $C<C^{*}$, two additional fixed points are observed to appear at the coordinates $\left(\rho_{A}^{\pm},1-\rho_{A}^{\pm}\right)$ where 
\begin{equation}\label{rhoaTransition}
    \rho_{A}^{\pm} = \frac{1}{2}\left(1\pm\sqrt{1+\frac{4C}{C-\frac{3N^{2}}{2N+C}}}\right).
\end{equation}
 Furthermore, they are found to be stable for $C<C^{t}$ and unstable otherwise, where $C^{t}\approx 0.1547N$. $C^{t}$ denotes the \textit{transition} value, as it predicts the occurrence of a transition in the hypergraph resulting in metastable state formation. Similar behaviour was observed by Lambiotte et.\ al \cite{lambiotte2007majority} in their study of MR dynamics on modular networks. As $C$ tends to $0$ the metastable state coordinates converge to $\left(0,1\right)$ and $\left(1,0\right)$. While these points are asymptotically stable in the phase plane for $C<C^{t}$, uniform consensus will ultimately be reached. This is due to the fact that diffusion governs the evolution of the system once these metastable states are reached. 
 
 Figure \ref{PhaseDiagram} plots the phase diagram for the symmetric modular hypergraph. The connectivity parameter is varied along the horizontal axis, and the absolute difference between the density of states at stationarity is plotted on the vertical axis. Asymmetric initial conditions were assumed, with $\rho_{A}\left(0\right)=1$ and $\rho_{B}\left(0\right)=0$. The theoretical transition value $C^{t}$ is indicated by the dashed line, and is in very good agreement with the simulation results.

\subsection{The Asymmetric Case}
We extend our analysis by considering the scenario in which $p_{30}\neq p_{03}$ and $p_{21}\neq p_{12}$ in general, hence leading to an asymmetry in the hypergraph structure. The parameter $C_{AB}\in\left[0,N\right]$ dictates the distribution of $2N^{3}/3$ hyperedges amongst the $N_{30}$ and $N_{21}$ categories, whereas the parameter $C_{BA}\in\left[0,N\right]$ distributes $2N^{3}/3$ hyperedges amongst the $N_{03}$ and $N_{12}$ categories. The hyperedge selection parameters are given by
\begin{align}
    p_{30}&=\frac{1}{2}\left(\frac{N}{N+C_{AB}}\right)^{2},\label{p30Asym} \\ p_{21}&=\frac{NC_{AB}}{\left(N+C_{AB}\right)^{2}}+\frac{1}{2}\left(\frac{C_{AB}}{N+C_{AB}}\right)^{2},\label{p21Asym} \\
     p_{12}&=\frac{NC_{BA}}{\left(N+C_{BA}\right)^{2}}+\frac{1}{2}\left(\frac{C_{BA}}{N+C_{BA}}\right)^{2},\label{P12Asym}\\
     p_{03}&=\frac{1}{2}\left(\frac{N}{N+C_{BA}}\right)^{2}. \label{p03Asym}
\end{align}
Using these probabilities, the hyperedge distribution may be determined as a function of the interconnectivity parameters:
\begin{align}
    &N_{30} = \frac{2N^{5}}{3\left(N+C_{AB}\right)^{2}}, \label{AsymN30}\\
    &N_{21} = \frac{4N^{3}C_{AB}}{3\left(N+C_{AB}\right)^{2}}\left(N+\frac{C_{AB}}{2}\right),\label{AsymN21} \\
    &N_{12} = \frac{4N^{3}C_{BA}}{3\left(N+C_{BA}\right)^{2}}\left(N+\frac{C_{BA}}{2}\right), \label{AsymN12} \\
    &N_{03} = \frac{2N^{5}}{3\left(N+C_{BA}\right)^{2}}. \label{AsymN03}
\end{align}
Simulations are conducted using the hyperedge selection probabilities given in equations (\ref{p30Asym})-(\ref{p03Asym}), which is equivalent to sampling uniformly at random from the hyperedge distribution given in equations (\ref{AsymN30})-(\ref{AsymN03}). The drift velocities $v_{A}$ and $v_{B}$ may be calculated in the usual way, and are found to be $O\left(1\right)$ in all parameter regimes (explicit expressions may be found in Appendix \ref{AsymmetricmodularAppendix}). The diffusion terms $D_{A}$ and $D_{B}$ are found to be $O\left(1/N\right)$ for all $C$, implying that the drift velocities dominate the dynamics of the system for large $N$.
 In this case the deterministic model consists of the system $d\rho_{A}/dt=v_{A}$ and $d\rho_{B}/dt = v_{B}$, with $v_{A}$ and $v_{B}$ given in equations (\ref{vaAsymmetric}) and (\ref{vbAsymmetric}) respectively.
The fixed points $\left(0,0\right)$ and $\left(1,1\right)$ are asymptotically stable. Linear stability analysis at the fixed point $\left(1/2,1/2\right)$ is conducted numerically. Visual inspection of the vector fields associated with this system indicates that in certain parameter regimes, stable fixed points emerge in the vicinity of the points $\left(0,1\right)$ and $\left(1,0\right)$ in the phase plane. In the symmetric case, these fixed points were found to lie on the line $\rho_{B}=1-\rho_{A}$. However, in the asymmetric case this is no longer true. As $C_{AB}$ and $C_{BA}$ are increased from $0$ these points are observed to move away from the line $\rho_{B}=1-\rho_{A}$. Despite this, the absorbing states are positioned close to this line, which allows us to determine their domain of existence heuristically. The drift velocities $d\rho_{A}/dt$ and $d\rho_{B}/dt$ are denoted by the functions $f\left(\rho_{A},\rho_{B}\right)$ and $g\left(\rho_{A},\rho_{B}\right)$ respectively. 

The fixed point coordinates are approximated by solving the following system:
\begin{align}
    &f\left(\rho_{A},1-\rho_{A}\right) = 0, \label{frhoa}\\ &g\left(\rho_{A},1-\rho_{A}\right) = 0. \label{grhoa}
\end{align}
Solving equation (\ref{frhoa}) yields two solutions for $\rho_{A}$, referred to as $\rho_{f}^{\pm}$. Similarly, solving equation (\ref{grhoa}) yields two solutions for $\rho_{A}$, referred to as $\rho_{g}^{\pm}$. Explicit expressions for $\rho_{f}^{\pm}$ and $\rho_{g}^{\pm}$ are given in Appendix \ref{AsymmetricmodularAppendix}.
%\begin{align*}
%    \rho_{f}^{\pm} = \frac{1}{2}\left(1\pm\sqrt{1+\frac{4\left(\frac{2NC_{BA}}{\left(N+C_{BA}\right)^{2}}+\left(\frac{C_{AB}}{N+C_{AB}}\right)^{2}\right)}{\frac{C_{AB}\left(4N-C_{AB}\right)}{\left(N+C_{AB}\right)^{2}}+2\frac{C_{BA}\left(C_{BA}-N\right)}{\left(N+C_{BA}\right)^{2}}-3\frac{N^{2}}{\left(N+C_{AB}\right)^{2}}}}\right).
%\end{align*}
%Similarly, solving equation (\ref{grhoa}) yields the following solutions for $\rho_{A}$ which we shall refer to as $\rho_{g}^{\pm}$:
%\begin{align*}
%    \rho_{g}^{\pm} = \frac{1}{2}\left(1\pm\sqrt{1+\frac{4\left(\frac{2NC_{AB}}{\left(N+C_{AB}\right)^{2}}+\left(\frac{C_{BA}}{N+C_{BA}}\right)^{2}\right)}{\frac{C_{BA}\left(4N-C_{BA}\right)}{\left(N+C_{BA}\right)^{2}}+2\frac{C_{AB}\left(C_{AB}-N\right)}{\left(N+C_{AB}\right)^{2}}-3\frac{N^{2}}{\left(N+C_{BA}\right)^{2}}}}\right).
%\end{align*}
When $C_{AB}=C_{BA}$, $\rho_{f}^{\pm}$ and $\rho_{g}^{\pm}$ reduce to the expressions for $\rho_{A}^{\pm}$ given by equation (\ref{rhoaTransition}), where a correspondence between signs is understood. When $C_{AB}=C_{BA}=0$, $\rho_{f}^{+}=\rho_{g}^{+}=1$ and $\rho_{f}^{-}=\rho_{g}^{-}=0$. By continuity,  $\rho_{f}^{+}$ and $\rho_{g}^{+}$ are expected to be approximately equal for $C_{AB}=\epsilon_{1}$ and $C_{BA}=\epsilon_{2}$ where $\epsilon_{1}$ and $\epsilon_{2}$ are small positive perturbation parameters (and similarly for $\rho_{f}^{-}$ and $\rho_{g}^{-}$). However, once the approximation $\rho_{f}^{\pm}\approx\rho_{g}^{\pm}$ becomes invalid it follows that the system of equations (\ref{frhoa}) and (\ref{grhoa}) no longer has any fixed points. We proceed by defining a distance threshold $\varepsilon$ such that if  $\lVert\left(\rho_{f}^{\pm},1-\rho_{f}^{\pm}\right)-\left(\rho_{g}^{\pm},1-\rho_{g}^{\pm}\right)\rVert>\varepsilon$ then $\rho_{f}^{\pm}$ and $\rho_{g}^{\pm}$ are no longer deemed to be solutions to equations (\ref{frhoa}) and (\ref{grhoa}). Distance is calculated in the Euclidean norm. 

Using this threshold, a bifurcation diagram may be constructed numerically. For a given set of connectivity parameters, metastable state existence is inferred if the distance threshold criterion is satisfied, and if the eigenvalues of the Jacobian when evaluated at $\left(\rho_{f}^{\pm},1-\rho_{f}^{\pm}\right)$ and $\left(\rho_{g}^{\pm},1-\rho_{g}^{\pm}\right)$ have negative real parts. A numerical exploration of the parameter space indicates that $\varepsilon = 0.115$ is a sensible value for the threshold parameter. 
 \begin{figure}
    \centering
    \includegraphics[width=0.9\linewidth,height=0.7\linewidth]{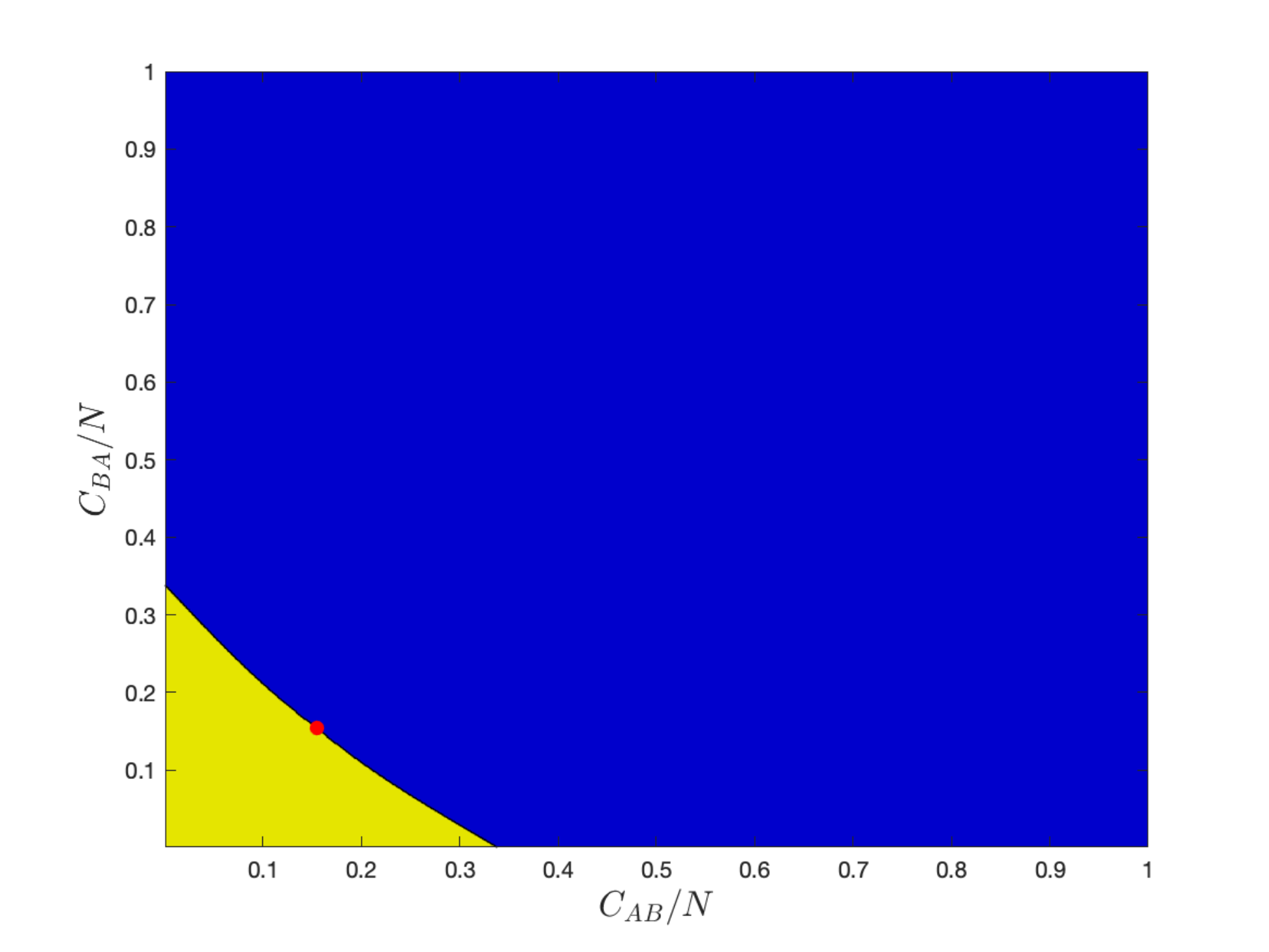}
    \begin{tikzpicture}[style={align=center}]
\footnotesize
\path [fill=myyellow] (0,0) rectangle (0.5,0.5);
    \node [right] at (0.5,0.2) {Stable} ;
\path [fill=myblue] (2.3,0) rectangle (2.8,0.5);
    \node [right] at (2.9,0.2) {Unstable/Nonexistent};
\end{tikzpicture}
    \caption[Bifurcation diagram for the asymmetric modular hypergraph]{Metastable state bifurcation plot for the asymmetric modular hypergraph.}
    \label{AsymmetricBifurcations}
\end{figure}
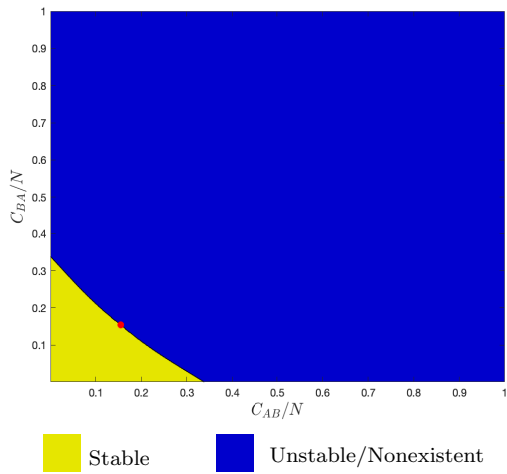
 The resulting bifurcation plot for the metastable states is given in Figure \ref{AsymmetricBifurcations}. The plot is symmetric about the line $C_{AB}=C_{BA}$. The red marker is located at $\left(C^{t}/N,C^{t}/N\right)$ where $C^{t}=0.1547N$ is the transition value on the symmetric modular hypergraph. The marker lies on the bifurcation boundary, as expected. Testing the nature of parameter coordinates close to the boundary indicates that the region of stability in the bifurcation plane is generally a very good predictor of the true region of stability in parameter space. \newline
\begin{figure}
    \centering
    \includegraphics[width=0.9\linewidth,height=0.7\linewidth]{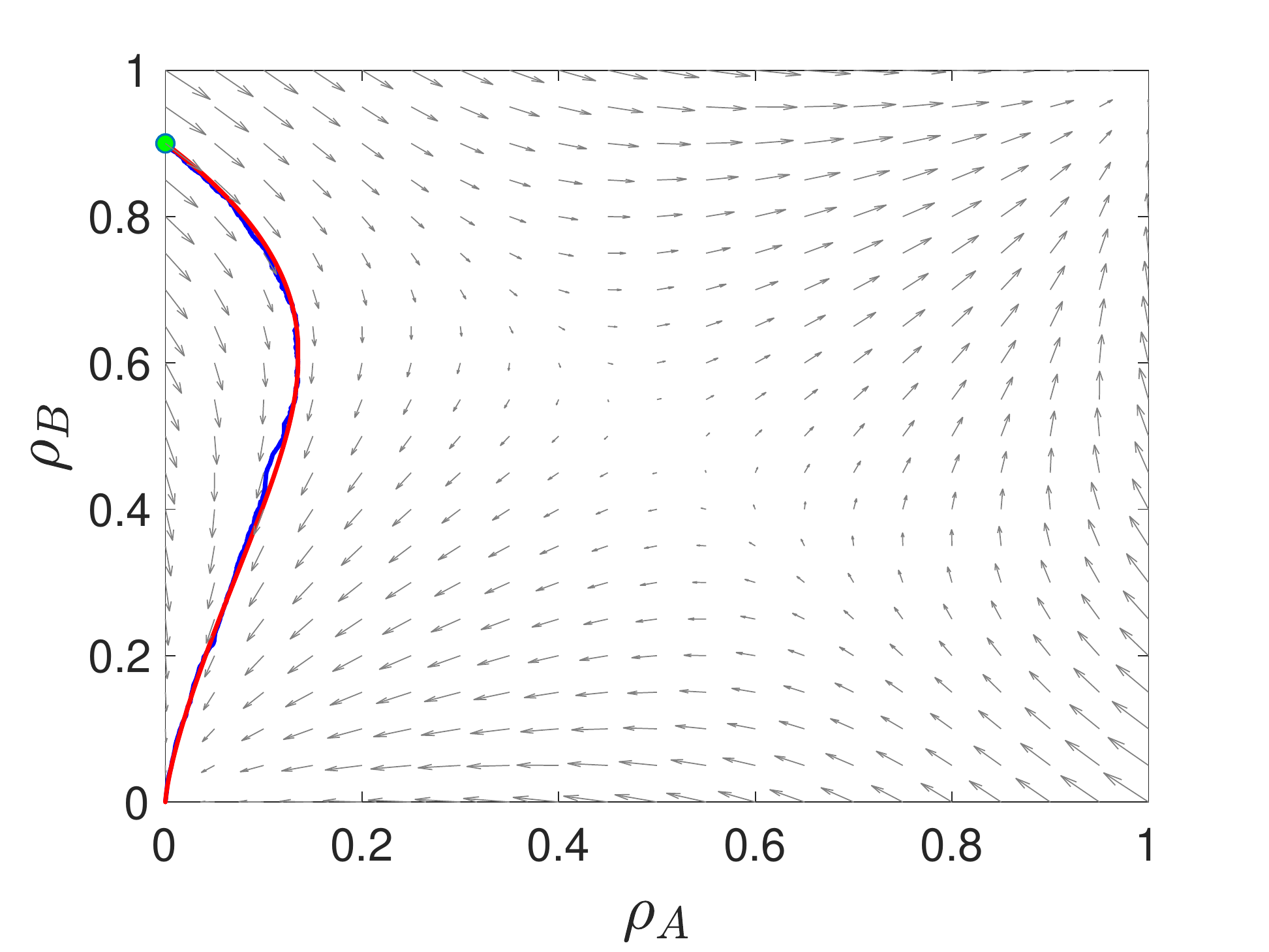}
    \caption{Simulations on the asymmetric modular hypergraph with $N=10^{4}$, $\boldsymbol\rho_{0}=\left(0,0.9\right)$ and $\left(C_{AB},C_{BA}\right)=\left(0.2N,0.7N\right)$.}
    \label{Asym27}
\end{figure}
%\begin{figure}
%     \centering
%     \begin{subfigure}[b]{0.45\linewidth}
%         \centering
%         \includegraphics[width=\linewidth,height=\linewidth]{C98ASYMLargeAXES.eps}
%         \caption{$\left(C_{AB},C_{BA}\right)=\left(0.9N,0.8N\right)$}
%         \label{AsymC98}
%     \end{subfigure}
%     \hfill
%     \begin{subfigure}[b]{0.45\linewidth}
%         \centering
%         \includegraphics[width=\linewidth,height=\linewidth]{C84ASYMLargeAXES.eps}
%         \caption{$\left(C_{AB},C_{BA}\right)=\left(0.8N,0.4N\right)$}
%         \label{AsymC57}
%     \end{subfigure}
%     \hfill
%     \begin{subfigure}[b]{0.45\linewidth}
%         \centering
%         \includegraphics[width=\linewidth,height=\linewidth]{C27ASYMLargeAXES.eps}
%         \caption{$\left(C_{AB},C_{BA}\right)=\left(0.2N,0.7N\right)$}
%         \label{AsymC32}
%     \end{subfigure}
%        \caption[Simulations on the asymmetric modular hypergraph]{Simulations on the asymmetric modular hypergraph with $N=10^{4}$ and $\boldsymbol\rho_{0}=\left(0,0.9\right)$.}
        \label{AsymmetricHighC}
%\end{figure}
\noindent Figure \ref{Asym27} provides simulation results on the asymmetric modular hypergraph with $\left(C_{AB},C_{BA}\right)=\left(0.2N,0.7N\right)$. Figure \ref{AsymmetricBifurcations} confirms that metastable states do not emerge in this instance. The red deterministic trajectory  is an excellent predictor of the blue stochastic path. This analysis indicates that the exit probability is governed by the geometry of the deterministic phase plane in the asymptotic limit as $N\rightarrow\infty$ for parameter regimes in which metastable states do not emerge. 

Diffusion becomes the dominating factor in dictating how the system reaches consensus for parameter coordinates in the stable region of Figure \ref{AsymmetricBifurcations}, due to metastable state formation. These metastable states tend to the points $\left(1,0\right)$ and $\left(0,1\right)$ as $C_{AB}$ and $C_{BA}$ tend to zero, illustrating that local consensus is reached independently in each community. 
\section{Generalised Heterogeneous Hypergraphs}\label{HeterogeneousNetworks}
\subsection{Heterogeneous Mean Field Analysis}
In this Section, we consider the scenario where the hyperedge distribution of a hypergraph $\mathcal{H}$ is given by a prescribed degree distribution, which we model with a heterogeneous mean field approximation \cite{landry2020effect}, as is often done in network science. Consider a system composed of $N$ nodes with a given degree distribution. We define the associated hypergraph $\mathcal{H}$ to consist of the node set $V\left(\mathcal{H}\right) = \{1,\dots,N\}$ and the set of triangles $T\left(\mathcal{H}\right) = \{\{i,j,k\}:i,j,k\in V\left(\mathcal{H}\right)\}$. As noted by Neuhauser et al.\ \cite{neuhauser2020opinion}, the structure of the hypergraph can be encoded in the adjacency tensor $\textbf{A}\in \mathbb{R}^{N\times N\times N}$ with entries 
\begin{equation}
    \textbf{A}_{ijk} = 
    \begin{cases}
    1 \quad \text{if}\quad \{i,j,k\}\in T\left(\mathcal{H}\right), \\
    0 \quad \text{otherwise}.
    \end{cases}
\end{equation}
The adjacency tensor is symmetric in all dimensions. In order to prevent the same node from being chosen multiple times in the formation of a triangle, we assume that $\textbf{A}_{ijk}=0$ if any two of the indices $i,j,k$ are the same. As usual, each node is assumed to occupy of two states, denoted by $\textbf{0}$ and $\textbf{1}$. The degree distribution $n_{k}$ is given by 
\begin{equation}\label{nk}
    n_{k} = \frac{N_{k}}{N},
\end{equation}
where $N_{k}$ is the number of nodes of degree $k$. The moments of the degree distribution are given by 
\begin{equation}\label{mu}
    \mu_{m} = \sum_{k}k^{m}n_{k}.
\end{equation}
Note that $\mu_{1}$ corresponds to the average node degree $\left<k\right>$. Let us denote the entire state of the system by $\eta$ and define $\eta\left(x\right)$ as the state of node $x$, which can take the values $0$ and $1$ (representative of the $\textbf{0}$ and $\textbf{1}$ states respectively). During each update event, exactly one node can change state. We represent the state of the system after changing the state of node $x$ by $\eta_{x}$, where
\begin{equation}
    \eta_{x} = 
    \begin{cases}
    \eta\left(y\right)\quad\text{if}\quad y\neq x, \\
    1-\eta\left(x\right)\quad\text{if}\quad y=x.
    \end{cases}
\end{equation}
We first consider a fully-connected hypergraph. That is, any three nodes can be chosen to form an interacting triangle. Suppose we choose a node uniformly at random, which we shall refer to as node $x$. The probability of doing so is $1/N$. To form a triangle, we can choose any two other nodes, of which there are ${{N-1}\choose 2}$ ways of doing so. Therefore the transition probability at node $x$ can be calculated as follows:
\begin{equation}
    \mathbb{P}\left[\eta\rightarrow\eta_{x}\right] = \frac{1}{N}\frac{1}{{{N-1}\choose2}}\sum_{yz}\textbf{A}_{xyz}\xi\left(x,y,z\right),
\end{equation}
where the indices $y$ and $z$ sum over all of the $N-1$ remaining nodes in the hypergraph and $\xi\left(x,y,z\right)$ is an indicator-type function given as follows:
\begin{align}\label{eta}
    \xi\left(x,y,z\right) &= \eta\left(y\right)\eta\left(z\right)\left(1-\eta\left(x\right)\right) \nonumber \\ &+\left(1-\eta\left(y\right)\right)\left(1-\eta\left(z\right)\right)\eta\left(x\right).
\end{align}
Note that $\xi\left(x,y,z\right)$ is $1$ if node $x$ is in the minority opinion in the triangle $\{x,y,z\}$ and $0$ otherwise. The presence of the adjacency tensor entries $\textbf{A}_{xyz}$ as multiplying factors in the summation prevent the same node from being chosen twice during triangle formation. The first term on the right hand side of equation (\ref{eta}) ensures that node $x$ flips from state $\textbf{0}$ to state $\textbf{1}$ if $x$ is in state $\textbf{0}$ and nodes $y$ and $z$ are in state $\textbf{1}$. The second term on the right hand side ensures that node $x$ flips from state $\textbf{1}$ to state $\textbf{0}$ if $x$ is in state $\textbf{1}$ and nodes $y$ and $z$ are in state $\textbf{0}$. Let $\rho_{k}\left(t\right)$ denote the density of $\textbf{1}$'s on nodes of fixed degree $k$ at time $t$. We shall write $\rho_{k}\left(t\right)$ as $\rho_{k}$ for convenience. Furthermore let $\rho_{k}^{\pm}=\rho_{k}\pm \delta\rho_{k}$ where $\delta\rho_{k}=N_{k}^{-1}$.  Let $R_{k}\left[\{\rho_{k}\}\right]$ and $L_{k}\left[\{\rho_{k}\}\right]$ denote the raising and lowering operators associated with nodes of degree $k$. They are calculated as follows:
\begin{align}
    &R_{k}\left[\{\rho_{k}\}\right] = \mathbb{P}\left[\rho_{k}\rightarrow\rho_{k}^{+}\right] \nonumber \\ &= \frac{1}{N}\frac{1}{{{N-1}\choose2}}\sum_{yz}\sum_{x}^{}{}^{'}\textbf{A}_{xyz}\eta\left(y\right)\eta\left(z\right)\left(1-\eta\left(x\right)\right), \label{Rklong}\\
    &L_{k}\left[\{\rho_{k}\}\right] = \mathbb{P}\left[\rho_{k}\rightarrow\rho_{k}^{-}\right] \nonumber\\ &= \frac{1}{N}\frac{1}{{{N-1}\choose2}}\sum_{yz}\sum_{x}^{}{}^{'}\textbf{A}_{xyz}\left(1-\eta\left(y\right)\right)\left(1-\eta\left(z\right)\right)\eta\left(x\right),\label{Lklong}
\end{align}
where the primed summation indicates that the summation is restricted to nodes $x$ of degree $k$. We now appeal to the heterogeneous mean field approximation \cite{landry2020effect} in assuming that node states are independent and nodes of the same degree behave similarly. We proceed by replacing the entries of the adjacency tensor as follows:
\begin{equation}\label{HetMF}
    \textbf{A}_{xyz} \rightarrow \frac{2k_{x}k_{y}k_{z}}{N^{2}\mu_{1}^{2}}.
\end{equation}
Equation (\ref{HetMF}) gives the interaction probability between three nodes of degrees $k_{x},k_{y}$ and $k_{z}$, where the scaling constants are chosen for the purpose of normalization \cite{landry2020effect}. This results in triangles being clustered around nodes of higher degrees, which is more realistic than a uniform random distribution of triangles. Finally, let us define the degree-weighted moments $\omega_{m}$ \cite{sood2008voter} of the degree distribution:
\begin{equation}\label{omega}
    \omega_{m} = \frac{1}{N\mu_{m}}\sum_{x}k_{x}^{m}\eta\left(x\right) = \frac{1}{\mu_{m}}\sum_{k}k^{m}n_{k}\rho_{k}.
\end{equation}
Using equations (\ref{HetMF}) and (\ref{omega}), equations (\ref{Rklong}) and (\ref{Lklong}) reduce to the following expressions for large $N$:
\begin{align}
    &R_{k}\left[\{\rho_{k}\}\right] 
     \approx \frac{4kn_{k}\omega_{1}^{2}\left(1-\rho_{k}\right)}{N^{2}}, \label{Rk} \\
    &L_{k}\left[\{\rho_{k}\}\right] \approx \frac{4kn_{k}\rho_{k}\left(1-\omega_{1}\right)^{2}}{N^{2}}. \label{Lk}
\end{align}
%Here we note that $k_{x} = k$ is invariant under the primed summation and we use the fact that $\sum_{y}k_{y} = \sum_{z}k_{z}\approx N\mu_{1}$, seeing as the indices $y$ and $z$ sum over all but one of the nodes in the hypergraph. Finally we note that for sufficiently large $N$, $ N^{2}/\left(\left(N-1\right)\left(N-2\right)\right)\approx 1$. \newline

\noindent Let $p\left(\{\rho_{k}\},t\right)$ denote the probability that the density of $\textbf{1}$'s on nodes of degree $k$ is $\rho_{k}$ at time $t$. We denote $p\left(\{\rho_{k}\},t\right)$ by $p$ for ease of notation, and similarly for $R_{k}$ and $L_{k}$. The probability density obeys the following master equation:
\begin{align}
    p\left(t+\delta t\right) &= \sum_{k}R_{k}\left(\rho_{k}^{-}\right)p\left(\rho_{k}^{-},t\right) + \sum_{k}L_{k}\left(\rho_{k}^{+}\right)p\left(\rho_{k}^{+},t\right) \nonumber \\
    &+\left[1-\sum_{k}\left(R_{k}+L_{k}\right)\right]p.
    \label{masterk}
\end{align}
Taylor expanding equation (\ref{masterk}) yields the following Fokker-Planck equation:
\begin{align}
    \frac{\partial p}{\partial t} &= -\sum_{k}\frac{\delta\rho_{k}}{\delta t}\frac{\partial }{\partial\rho_{k}}\left(\left(R_{k}-L_{k}\right)p\right)\nonumber \\&+\sum_{k}\frac{\left(\delta\rho_{k}\right)^{2}}{2\delta t}\frac{\partial^{2}}{\partial\rho_{k}^{2}}\left(\left(R_{k}+L_{k}\right)p\right).\label{FokkerPlankKDensities}
\end{align}
Using equation (\ref{FokkerPlankKDensities}) the drift velocities $\{v_{k}\}$ and diffusion terms $\{D_{k}\}$ of the densities $\{\rho_{k}\}$ may be identified:
\begin{align}
    &v_{k} = \frac{\delta\rho_{k}}{\delta t}\left(R_{k}-L_{k}\right) \nonumber \\ &= \frac{4k}{N^{2}}\left(\left(1-\rho_{k}\right)\omega_{1}^{2}-\rho_{k}\left(1-\omega_{1}\right)^{2}\right), \label{vk} \\
    &D_{k} = \frac{\left(\delta\rho_{k}\right)^{2}}{2\delta t}\left(R_{k}+L_{k}\right) \nonumber \\  &= \frac{2k}{N^{2}N_{k}}\left(\left(1-\rho_{k}\right)\omega_{1}^{2}+\rho_{k}\left(1-\omega_{1}\right)^{2}\right) \label{Dk}.
\end{align}
From equations (\ref{vk}) and (\ref{Dk}) we observe $D_{k}$ is suppressed by a factor of $1/N_{k}$ relative to $v_{k}$. This suggests that the dynamics of the densities $\{\rho_{k}\}$ are dominated by the drift velocities $\{v_{k}\}$ when $N_{k}$ is large. Under this assumption, we posit that the dynamics of the system may be well described by the set of equations
\begin{equation}\label{rhoksystem}
    \frac{d\rho_{k}}{dt} = \frac{4k}{N^{2}}\left(\left(1-\rho_{k}\right)\omega_{1}^{2}-\rho_{k}\left(1-\omega_{1}\right)^{2}\right)
\end{equation}
 where $1\leq k \leq k_{max}$ and $k_{max}$ is the maximum node degree in the network. The number of equations in this system is equal to the number of distinct node degrees. By symmetry, the system has fixed points when  all of the $\rho_{k}$'s are equal. Suppose $\rho_{k}=\rho'$ for all $k$, where $\rho'\in\left[0,1\right]$. Recalling the definition of $\omega_{1}$ from equation (\ref{omega}) we observe that $\omega_{1} = \rho'$ and therefore fixed points occur for $\rho'\in\{0,1/2,1\}$. The two states of uniform consensus are asymptotically stable. When $\rho_{k}=1/2$ for all $k$, linear stability analysis is conducted numerically.
 
 \subsection{Numerical Implementation}\label{3degree}
\begin{figure*}
     \centering
     \begin{subfigure}[b]{0.3\linewidth}
         \centering
         \includegraphics[width=\linewidth,height=\linewidth]{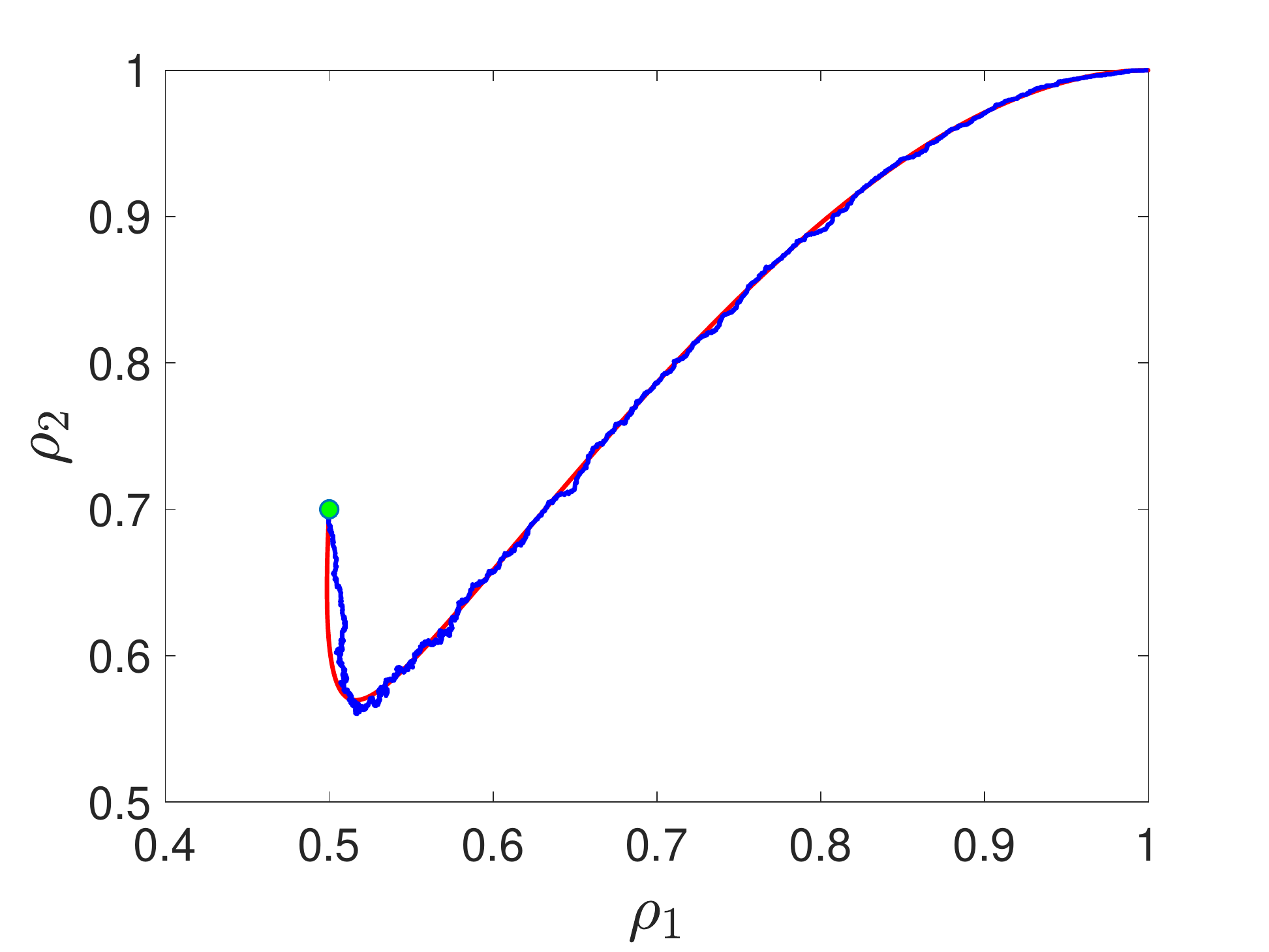}
         \caption{$\left(\rho_{1},\rho_{2}\right)$}
         \label{P1P2}
     \end{subfigure}
     \hfill
     \begin{subfigure}[b]{0.3\linewidth}
         \centering
         \includegraphics[width=\linewidth,height=\linewidth]{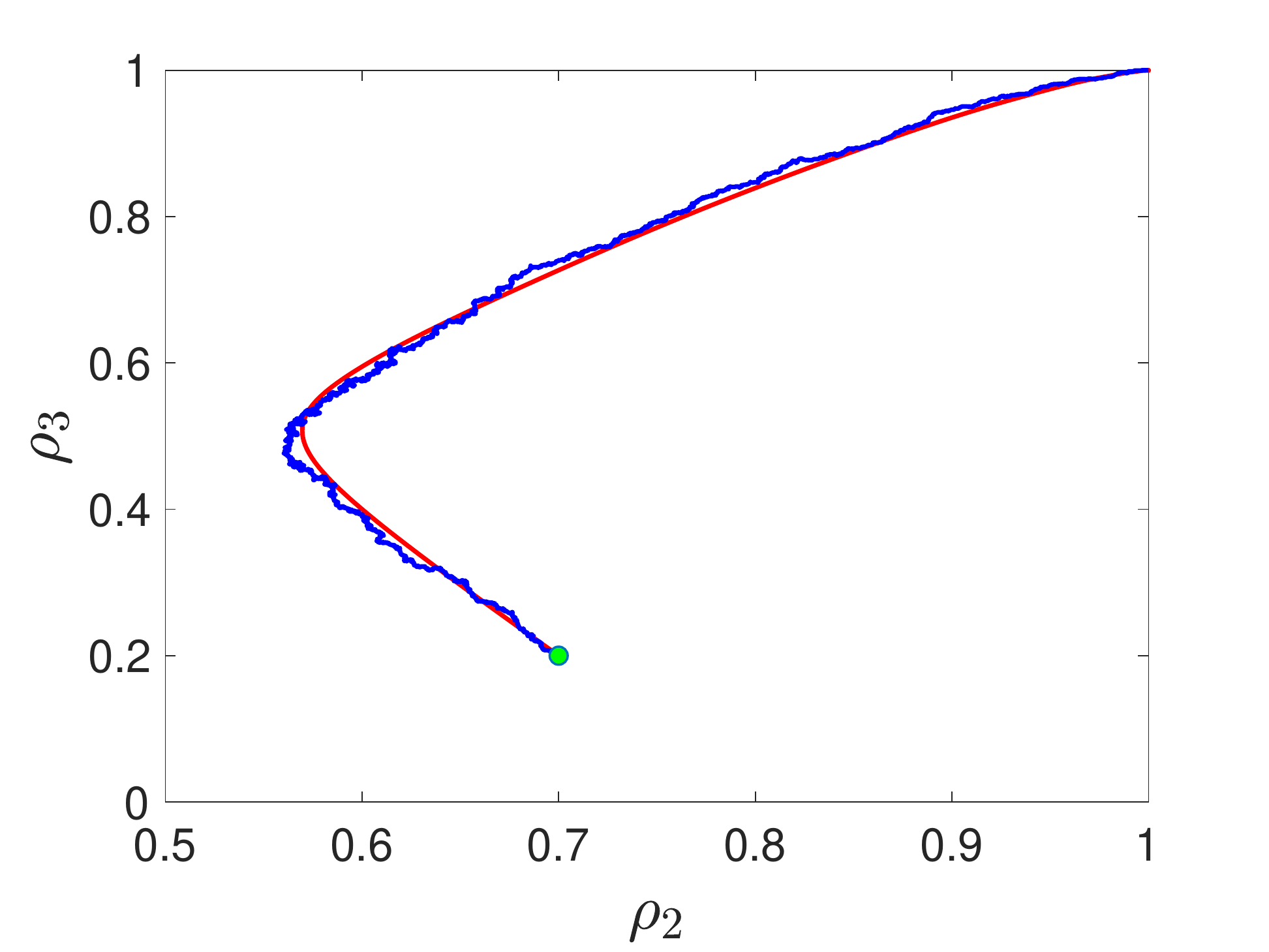}
         \caption{$\left(\rho_{2},\rho_{3}\right)$}
         \label{P2P3}
     \end{subfigure}
     \hfill
     \begin{subfigure}[b]{0.3\linewidth}
         \centering
         \includegraphics[width=\linewidth,height=\linewidth]{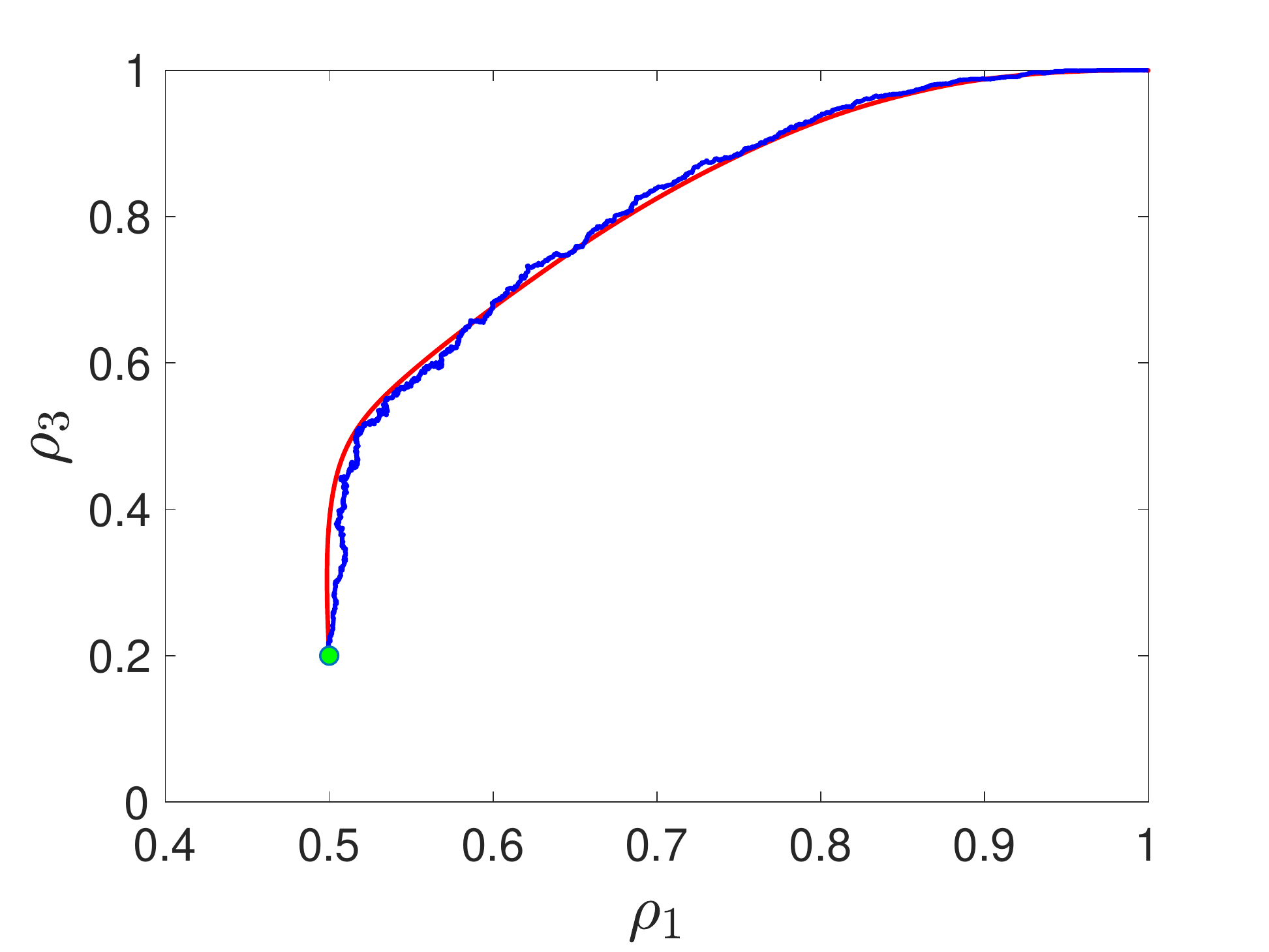}
         \caption{$\left(\rho_{1},\rho_{3}\right)$}
         \label{P1P3}
     \end{subfigure}
        \caption[Simulations on a simple heterogeneous network]{Two-dimensional phase plane projections associated with the heterogeneous network described in Section \ref{3degree}. In this instance, $N=1.2\times10^{4}$.}
        \label{Projections1}
\end{figure*}

  In this illustrative example, we consider a system where we impose that $1/2$ of the nodes have degree $1$, $1/3$ of the nodes have degree $2$ and $1/6$ of the nodes have degree $3$. Using equation (\ref{rhoksystem}) the associated dynamical system may be written as follows:
 \begin{align*}
     \frac{d\rho_{1}}{dt} = \frac{4}{N^{2}}\left(\omega_{1}^{2}\left(1-\rho_{1}\right)-\rho_{1}\left(1-\omega_{1}\right)^{2}\right), \\
     \frac{d\rho_{2}}{dt} = \frac{8}{N^{2}}\left(\omega_{1}^{2}\left(1-\rho_{2}\right)-\rho_{2}\left(1-\omega_{1}\right)^{2}\right), \\
     \frac{d\rho_{3}}{dt} = \frac{12}{N^{2}}\left(\omega_{1}^{2}\left(1-\rho_{3}\right)-\rho_{3}\left(1-\omega_{1}\right)^{2}\right). 
 \end{align*}
 Numerical analysis of the Jacobian reveals that $\left(1/2,1/2,1/2\right)$ is a saddle point. Figure \ref{Projections1} shows a sample set of simulation results on such a network, where the initial conditions are given by $\left(\rho_{1}\left(0\right),\rho_{2}\left(0\right),\rho_{3}\left(0\right)\right) = \left(0.5,0.7,0.2\right)$. Phase plane projections are provided given that the system is three-dimensional. The deterministic path is plotted in red and the stochastic path is plotted in blue. The two curves are practically superimposed which indicates that the deterministic model is generally an excellent predictor of the behaviour of the system. Similar results were observed for other choices of initial conditions. This indicates that the exit probability of the system is determined by the geometry of the deterministic phase plane as $N$ becomes large. 

\section{Conclusion}\label{Conclusion}
In this paper, we have argued that hypergraphs provide a natural, and efficient, framework to explore the relations between structure and dynamics in situations where basic interaction units involve more than two nodes. We have conducted an in-depth study of a generic model from opinion dynamics, the Majority Rule (MR), and have proposed
a number of  hypergraph models to study MR dynamics. Our analysis was achieved my recasting the dynamics in terms of Fokker-Planck equations. By deriving the Fokker-Planck equation associated with MR dynamics on a given hypergraph, the drift and diffusion terms governing the evolution of the system could be deduced. Interestingly, it was found that diffusive contributions to the stochastic dynamics vanished as the population size became increasingly large. This allowed for the system to be modelled using deterministic nonlinear dynamical systems which, in essence, reflected the non-conservative nature of MR. On all of the hypergraph topologies considered, the deterministic modelling approach proved to be an excellent predictor of the behaviour of the system, which allowed for the final state of consensus to be predicted as a function of the initial conditions. This is markedly different to the behaviour observed for VM dynamics on heterogeneous networks \cite{sood2008voter}, where diffusion is non-negligible even in large populations.

There are a number of ways in which the analysis presented in this paper could be extended. It would be desirable to find a way in which to accurately model MR dynamics on small hypergraphs. In this case, making analytical progress could be difficult as diffusion would be non-negligible, and as mean-field approximations would also be expected to be less relevant. However, a numerical study of the stochastic differential equations derived from the associated Fokker-Planck equations could still yield powerful insights. It would also be desirable to find a way in which to apply the analysis presented in Section \ref{HeterogeneousNetworks} to more sophisticated underlying network structures, as encoded by their adjacency tensor, or including interactions of arbitrary size.
%Efforts were made to implement the deterministic model on hypergraphs derived from underlying scale-free networks, though this proved challenging. Each time a scale-free network is generated via preferential attachment, there is no way of knowing in advance how many distinct node degrees will be present. This makes it difficult to implement the system given by equation $\left(\ref{rhoksystem}\right)$ computationally. 
Finally, it would be interesting to find a way in which to characterise the consensus time as a function of the population size. Krapivksy and Redner \cite{krapivsky2003dynamics} succeeded in doing so for the mean field case, though it remains to be investigated on more complex hypergraph topologies.

\begin{acknowledgments}
RL would like to thank Michael Schaub and Leonie Neuhauser for inspiring discussions related to this manuscript. 
\end{acknowledgments}
\onecolumngrid
\section{Linear Stability Analysis on the Symmetric Modular Hypergraph}
\label{modularAppendix}
Here we provide additional details of the analysis presented in Section \ref{SymmetricModularText}. We conduct linear stability analysis of the system
\begin{align*}
    d\rho_{A}/dt&=v_{A}, \\
    d\rho_{B}/dt&=v_{B}, \\
\end{align*}
 where $v_{A}$ and $v_{B}$ are given in equations (\ref{vaSymmetric}) and (\ref{vbSymmetric}) respectively. Let $\mathcal{J}$ denote the associated Jacobian matrix. The fixed points of the system are given by $\left(0,0\right),\left(1/2,1/2\right)$ and $\left(1,1\right)$. The points $\left(0,0\right)$ and $\left(1,1\right)$ are asymptotically stable. The analysis of the fixed point at $\left(1/2,1/2\right)$ is slightly more complicated. The matrix $\mathcal{J}\left(1/2,1/2\right)$ has eigenvalues $\lambda_{\pm}$ given by
\begin{equation*}
    \lambda_{\pm}=\frac{3}{2}\left(\frac{N}{N+C}\right)^{2}-\frac{C\left(2N+C\right)}{2\left(N+C\right)^{2}}\pm \frac{2C\left(2N+C\right)}{\left(N+C\right)^{2}}.
\end{equation*}
$\lambda_{+}$ is always positive, therefore $\left(1/2,1/2\right)$ is unstable for all $C$. For $C=0$ we have $\lambda_{-}>0$ and for $C=N$ we have $\lambda_{-}<0$. In order to determine the value of $C$ for which $\left(1/2,1/2\right)$ transitions between a saddle point and a source, we solve the equation $\lambda_{-}\left(C\right)=0$. There are two possible values of $C$ (denoted by $C_{\pm}$) for which this equation is satisfied:
\begin{equation*}
    C_{\pm} = \left(\frac{-5\pm2\sqrt{10}}{5}\right)N.
\end{equation*}
$C_{-}$ is negative, therefore it is nonphysical. Hence we conclude that the the fixed point at $\left(1/2,1/2\right)$ changes in nature when $C_{+} \equiv C^{*} \approx 0.2649N$. \newline

\noindent Visual inspection of the $\left(\rho_{A},\rho_{B}\right)$ phase plane reveals that two new unstable fixed points emerge along the line $\rho_{B} = 1 - \rho_{A}$ when $C$ becomes sufficiently small. Furthermore, numerical simulations indicate that for some critical value of $C$, which we shall refer to as $C^{t}$, the two new fixed points transition between being stable and unstable. That is to say for $C\in\left[0,C^{t}\right]$ there exist four absorbing states in the phase plane. In order to determine the coordinates of the two new fixed points, and the range of values of $C$ for which they exist, the drift velocities are evaluated along the line $\rho_{B}=1-\rho_{A}$. We do so by rewriting equations the dynamical system as $d\rho_{A}/dt = f\left(\rho_{A},\rho_{B}\right)$ and $d\rho_{B}/dt = g\left(\rho_{A},\rho_{B}\right)$ respectively where $f$ and $g$ are functions of the clique densities. By symmetry, it suffices to consider only one of these equations given that we are interested in seeking solutions to the equation $f\left(\rho_{A},1-\rho_{A}\right)=0$. This equation ultimately reduces to a quadratic in $\rho_{A}$, yielding the two solutions given in equation (\ref{rhoaTransition}):
\begin{equation*}
    \rho_{A}^{\pm} = \frac{1}{2}\left(1\pm\sqrt{1+\frac{4C}{C-\frac{3N^{2}}{2N+C}}}\right).
\end{equation*}
Using the expressions for $\rho_{A}^{\pm}$ in conjunction with the relation $\rho_{B}=1-\rho_{A}$ gives the coordinates of the two new fixed points. As $C$ tends to $0$ equation (\ref{rhoaTransition}) predicts that the fixed points tend to the coordinates $\left(0,1\right)$ and $\left(1,0\right)$. We can determine the range of values of $C$ for which these fixed points exist by looking at where the discriminant in equation (\ref{rhoaTransition}) is positive. Solving the resultant quadratic equation reveals that the fixed points exist on the interval $C\in\left[0,C^{*}\right]$. This illustrates that as we approach $C^{*}$ from above, the two new fixed points spontaneously emerge at the exact instant when the saddle point at $\left(1/2,1/2\right)$ becomes a source. In order to determine the stability of these two fixed points, once again we proceed using linear stability analysis. When evaluating the Jacobian along the line $\rho_{B}=1-\rho_{A}$ we find that $\mathcal{J}_{11}=\mathcal{J}_{22}$ and $\mathcal{J}_{12}=\mathcal{J}_{21}$, where $\mathcal{J}_{ij}$ denotes the Jacobian entry in row $i$ and column $j$. Therefore the eigenvalues are given by $\lambda_{\pm} = \mathcal{J}_{11}\pm\mathcal{J}_{12}$. Substituting $(\rho_{A}^{+},1-\rho_{A}^{+})$ into $\mathcal{J}$ yields the following expressions for the Jacobian entries:
\begin{equation*}
    \mathcal{J}_{11} = \frac{7C^{2}+14CN-3N^{2}}{\left(N+C\right)^{2}},\quad 
    \mathcal{J}_{12} = \frac{2C\left(2N+C\right)}{\left(N+C\right)^{2}}.
\end{equation*}
We note that $\mathcal{J}_{12}$ is always positive, therefore in order to determine the point at which these fixed points become stable it suffices to consider solutions to the equation $\mathcal{J}_{11}+\mathcal{J}_{12}=0$. This ultimately reduces to a quadratic equation in $C$, where the non-negative root $C^{t}$ is found to be
\begin{equation*}
    C^{t} = \left(\frac{-3+2\sqrt{3}}{3}\right)N\approx 0.1547N.
\end{equation*}
The fixed points are therefore stable for $C\in\left[0,C^{t}\right]$. In summary, $C^{t}$ is the parameter value below which metastable state formation is predicted by the deterministic model on the symmetric modular hypergraph. 
\onecolumngrid
\section{Explicit Parameter Expressions on the Asymmetric Modular Hypergraph}
\label{AsymmetricmodularAppendix}
Here we provide explicit expressions for $\rho_{f}^{\pm}$ and $\rho_{g}^{\pm}$, the solutions to equations (\ref{frhoa}) and (\ref{grhoa}) respectively:
\begin{align*}
    \rho_{f}^{\pm} &= \frac{1}{2}\left(1\pm\sqrt{1+\frac{4\left(\frac{2NC_{BA}}{\left(N+C_{BA}\right)^{2}}+\left(\frac{C_{AB}}{N+C_{AB}}\right)^{2}\right)}{\frac{C_{AB}\left(4N-C_{AB}\right)}{\left(N+C_{AB}\right)^{2}}+2\frac{C_{BA}\left(C_{BA}-N\right)}{\left(N+C_{BA}\right)^{2}}-3\frac{N^{2}}{\left(N+C_{AB}\right)^{2}}}}\right) \\
    \rho_{g}^{\pm} &= \frac{1}{2}\left(1\pm\sqrt{1+\frac{4\left(\frac{2NC_{AB}}{\left(N+C_{AB}\right)^{2}}+\left(\frac{C_{BA}}{N+C_{BA}}\right)^{2}\right)}{\frac{C_{BA}\left(4N-C_{BA}\right)}{\left(N+C_{BA}\right)^{2}}+2\frac{C_{AB}\left(C_{AB}-N\right)}{\left(N+C_{AB}\right)^{2}}-3\frac{N^{2}}{\left(N+C_{BA}\right)^{2}}}}\right).
\end{align*}

Expressions for the drift velocities on the asymmetric modular hypergraph are as follows:
\begin{align}
    &v_{A} = 3\left(\frac{N}{N+C_{AB}}\right)^{2}\rho_{A}\left(1-\rho_{A}\right)\left(2\rho_{A}-1\right) \nonumber \\ &+\frac{4NC_{AB}}{\left(N+C_{AB}\right)^{2}}\rho_{A}\left(1-\rho_{A}\right)\left(2\rho_{B}-1\right) \nonumber \\
    &+2\left(\frac{C_{BA}}{N+C_{BA}}\right)^{2}\rho_{A}\left(1-\rho_{A}\right)\left(2\rho_{B}-1\right) \nonumber \\
    &+\frac{2NC_{BA}}{\left(N+C_{BA}\right)^{2}}\left(\rho_{B}^{2}\left(1-\rho_{A}\right)-\rho_{A}\left(1-\rho_{B}\right)^{2}\right) \nonumber\\ &+\left(\frac{C_{AB}}{N+C_{AB}}\right)^{2}\left(\rho_{B}^{2}\left(1-\rho_{A}\right)-\rho_{A}\left(1-\rho_{B}\right)^{2}\right), \label{vaAsymmetric} \\
    &v_{B} = 3\left(\frac{N}{N+C_{BA}}\right)^{2}\rho_{B}\left(1-\rho_{B}\right)\left(2\rho_{B}-1\right) \nonumber \\ &+\frac{4NC_{BA}}{\left(N+C_{BA}\right)^{2}}\rho_{B}\left(1-\rho_{B}\right)\left(2\rho_{A}-1\right)\nonumber \\
    &+2\left(\frac{C_{AB}}{N+C_{AB}}\right)^{2}\rho_{B}\left(1-\rho_{B}\right)\left(2\rho_{A}-1\right) \nonumber \\
    &+\frac{2NC_{AB}}{\left(N+C_{AB}\right)^{2}}\left(\rho_{A}^{2}\left(1-\rho_{B}\right)-\rho_{B}\left(1-\rho_{A}\right)^{2}\right)\nonumber \\&+\left(\frac{C_{BA}}{N+C_{BA}}\right)^{2}\left(\rho_{A}^{2}\left(1-\rho_{B}\right)-\rho_{B}\left(1-\rho_{A}\right)^{2}\right). \label{vbAsymmetric}
\end{align}
Equations (\ref{vaAsymmetric}) and (\ref{vbAsymmetric}) reduce to equations (\ref{vaSymmetric}) and (\ref{vbSymmetric}) respectively when $C_{AB}=C_{BA}=C$.

\twocolumngrid
\bibliographystyle{apsrev4-1}
\bibliography{refs}        %use a bibtex bibliography file refs.bib

\end{document}